\documentclass[aps,prd,showpacs,notitlepage,nofootinbib,preprintnumbers,amsmath,amssymb]{revtex4-1}

\usepackage{graphics,graphicx}
\usepackage{dcolumn}
\usepackage{bm}
\usepackage{epsfig}
\usepackage[usenames]{color}
\usepackage{hyperref} 
\usepackage{mathbbol}
\usepackage{epstopdf}
\usepackage{simplewick}
\usepackage[utf8]{inputenc} 
\usepackage{slashed}
\usepackage{stackrel}

\def\eq#1{{Eq.~(\ref{#1})}}

\newcommand{\ben}{\begin{eqnarray*}}
\newcommand{\een}{\end{eqnarray*}}
\newcommand{\un}[1]{\underline{#1}}

\newcommand{\pd}{\partial}

\newcommand{\tr}{\mbox{tr}}
\newcommand{\thalf}{\tfrac{1}{2}}

\newcommand{\stackeven}[2]{{{}_{\displaystyle{#1}}\atop\displaystyle{#2}}}

\newcommand{\gsim}{\stackeven{>}{\sim}}
\newcommand{\as}{\alpha_s}

\newcommand{\dhd}{{\textstyle d}
\lower.03ex\hbox{\kern-0.38em$^{\scriptstyle-}$}\kern-0.05em{}}
\newcommand{\dbar}{{\textstyle \delta}
\lower.03ex\hbox{\kern-0.38em$^{\scriptstyle-}$}\kern-0.05em{}}
\newcommand{\half}{{1\over 2}}

\newcommand{\bra}[1]{\left\langle #1 \right|}
\newcommand{\ket}[1]{\left| #1 \right\rangle}

\newcommand{\ul}[1]{\underline{#1}}

\newcommand{\braket}[2]{\left\langle #1 \right| \left. #2 \right\rangle}

\newcommand{\cc}{\mbox{c.c.}}

\begin{document}

\title{Orbital Angular Momentum at Small $x$}

\author{Yuri V. Kovchegov} 
         \email[Email: ]{kovchegov.1@osu.edu}
         \affiliation{Department of Physics, The Ohio State
           University, Columbus, OH 43210, USA}

\begin{abstract}
  We determine the small Bjorken $x$ asymptotics of the quark and gluon orbital angular momentum (OAM) distributions in the proton in the double-logarithmic approximation (DLA), which resums powers of $\as \ln^2 (1/x)$ with $\as$ the strong coupling constant. Starting with the operator definitions for the quark and gluon OAM, we simplify them at small $x$, relating them, respectively, to the polarized dipole amplitudes for the quark and gluon helicities defined in our earlier works. Using the small-$x$ evolution equations derived for these polarized dipole amplitudes earlier we arrive at the following small-$x$ asymptotics of the quark and gluon OAM distributions in the large-$N_c$ limit:
  \begin{subequations}\label{OAM_abs}
  \begin{align}
 L_{q + \bar{q}} (x, Q^2) = - \Delta \Sigma (x, Q^2) \sim
  \left(\frac{1}{x}\right)^{\frac{4}{\sqrt{3}} \, \sqrt{\frac{\as
        \, N_c}{2 \pi}} },   \\
       L_G (x, Q^2) \sim \Delta G (x, Q^2) \sim 
  \left(\frac{1}{x}\right)^{\frac{13}{4 \sqrt{3}} \, \sqrt{\frac{\as
        \, N_c}{2 \pi}}} . 
\end{align}
\end{subequations}
\end{abstract}


\maketitle

\tableofcontents


%
\section{Introduction} \label{sec:Intro}
%

In recent years a lot of progress has been achieved in our theoretical understanding of quark and gluon helicity distributions at small $x$ \cite{Kovchegov:2015pbl,Kovchegov:2016weo,Kovchegov:2016zex,Kovchegov:2017jxc,Kovchegov:2017lsr,Kovchegov:2018znm}. While having precise control of helicity distributions is important for our understanding of the proton spin, another important component of the proton spin comes from the quark and gluon orbital angular momentum (OAM). The helicity sum
rules~\cite{Jaffe:1989jz,Ji:1996ek,Ji:2012sj} (see \cite{Leader:2013jra} for a review) include the quark and gluon contributions to the proton spin along with their OAM. Specifically, the Jaffe--Manohar sum rule~\cite{Jaffe:1989jz} reads
\begin{align}
  \label{eq:sum_rule}
  S_{q + \bar{q}} + L_{q + \bar{q}} + S_G + L_G = \frac{1}{2}\, .
\end{align}
Here 
\begin{align}
  \label{eq:net_spin}
  S_{q + \bar{q}} (Q^2) = \frac{1}{2} \, \int\limits_0^1 \!dx \, \Delta \Sigma (x,
  Q^2) \, \ \ \ \mbox{and} \ \ \ S_G (Q^2) = \int\limits_0^1 \!dx \,
  \Delta G (x, Q^2)\,,
\end{align}
are the quark and gluon components of the proton spin expressed in terms of the
quark and gluon helicity distributions $\Delta \Sigma (x, Q^2) = \sum_f \, \left[ \Delta q^f (x, Q^2)  +  \Delta \bar{q}^f (x, Q^2) \right]$ and $\Delta
G (x, Q^2)$. Importantly, $L_{q + \bar{q}}$ and $L_G$ in \eq{eq:sum_rule} denote the quark and gluon OAM, respectively. Our understanding of the proton spin would be incomplete without a good quantitative understanding of $L_{q + \bar{q}}$ and $L_G$.

The quark and gluon OAM can be written down as integrals of their distributions in $x$ \cite{Bashinsky:1998if,Hagler:1998kg,Harindranath:1998ve,Hatta:2012cs,Ji:2012ba}
\begin{align}
\label{eq:net_OAM}
  L_{q + \bar{q}} (Q^2) = \int\limits_0^1 \!dx \, L_{q + \bar{q}} (x,
  Q^2) \, \ \ \ \mbox{and} \ \ \ L_G (Q^2) = \int\limits_0^1 \!dx \,
  L_G (x, Q^2)\, .
\end{align}
It now becomes apparent that, just like the quark and gluon helicities, both the quark OAM and the gluon OAM may receive contributions from the small-$x$ region. Even if the difficulty in experimentally measuring the gluon OAM is somehow surmounted, any given experiment can measure $L_{q + \bar{q}} (x, Q^2)$ and $L_G (x, Q^2)$ only down to some minimal value of $x = x_{min}$. No matter how small are the values of $x_{min}$ to be accessed in the future experiments, one always faces the question of constraining the $x < x_{min}$ region. It appears that a solid theoretical understanding  of $L_{q + \bar{q}} (x, Q^2)$ and $L_G (x, Q^2)$ at small $x$ is necessary to accomplish this goal. The aim of this paper is to theoretically determine the small-$x$ asymptotics of the quark and gluon OAM. 

Important first steps in this direction were taken in \cite{Hatta:2018itc}, where the one-loop $Q^2$ evolution equations for $L_{q + \bar{q}} (x, Q^2)$ and $L_G (x, Q^2)$ \cite{Hagler:1998kg,Martin:1998fe,Hatta:2016aoc} were solved both numerically and analytically, with the aim of determining the small-$x$ asymptotics of these quantities. However, $Q^2$ evolution equations at the one-loop level resum powers of $\as \ln Q^2/Q_0^2$ with some initial momentum scale $Q_0$. At small $x$ these equations correctly reproduce the powers of  $\as \ln Q^2/Q_0^2 \, \ln (1/x)$. That is, they give one the small-$x$ asymptotics only at large values of $Q^2$. For spin-independent observables, the true small-$x$ asymptotics is obtained by resumming powers of $\as \ln (1/x)$ (without any ordering of the transverse momenta): this is done in the Balitsky--Fadin--Kuraev--Lipatov (BFKL)
\cite{Kuraev:1977fs,Balitsky:1978ic}, Balitsky--Kovchegov (BK)
\cite{Balitsky:1995ub,Balitsky:1998ya,Kovchegov:1999yj,Kovchegov:1999ua}
and Jalilian-Marian--Iancu--McLerran--Weigert--Leonidov--Kovner
(JIMWLK)
\cite{Jalilian-Marian:1997dw,Jalilian-Marian:1997gr,Iancu:2001ad,Iancu:2000hn}
evolution equations. For many spin-dependent observables, the leading small-$x$ contribution comes from resumming the powers of $\as \ln^2 (1/x)$ \cite{Kirschner:1983di, Kirschner:1985cb,
  Kirschner:1994vc,Kirschner:1994rq,Griffiths:1999dj,Itakura:2003jp,Bartels:2003dj} (such parameter does not exist in the unpolarized evolution).
This is the double-logarithmic approximation (DLA). In the case of helicity, the resummation of $\as \ln^2 (1/x)$ was addressed in \cite{Bartels:1995iu,Bartels:1996wc,Kovchegov:2015pbl,Kovchegov:2016weo,Kovchegov:2016zex,Kovchegov:2017jxc,Kovchegov:2017lsr,Kovchegov:2018znm}. Our aim here is to resum powers of $\as \ln^2 (1/x)$ for $L_{q + \bar{q}} (x, Q^2)$ and $L_G (x, Q^2)$, obtaining the leading small-$x$ asymptotics for these important quantities. To achieve this goal we will build upon our prior experience with the gluon helicity at small $x$ \cite{Kovchegov:2015pbl,Kovchegov:2016weo,Kovchegov:2016zex,Kovchegov:2017jxc,Kovchegov:2017lsr,Kovchegov:2018znm}.

Below we will start by analyzing the quark OAM at small $x$ in Sec.~\ref{sec:quarkOAM}.  We will first define the quark OAM operator using the quark Wigner distribution \cite{Belitsky:2003nz} in Sec.~\ref{sec:quarkOAMoperator} following \cite{Lorce:2011kd,Hatta:2011ku,Lorce:2011ni}. We then employ the technique from \cite{Kovchegov:2017lsr,Kovchegov:2018znm} to simplify the quark OAM operator at small $x$ in Sec.~\ref{sec:quarkOAMeval}. This leads to a relationship between $L_{q + \bar{q}} (x, Q^2)$ and the fundamental polarized dipole amplitude \cite{Kovchegov:2015pbl}, which is an expectation value of the polarized fundamental dipole operator defined in \cite{Kovchegov:2018znm}. While the quark helicity distribution was related to the impact-parameter integrated fundamental polarized dipole amplitude \cite{Kovchegov:2015pbl}, the quark OAM is related to the first impact parameter moment of this amplitude, as defined in Eqs.~\eqref{quarkOAM_moments}. Using the evolution equations constructed for the fundamental polarized dipole amplitude in \cite{Kovchegov:2015pbl,Kovchegov:2018znm}, in Sec.~\ref{sec:quarkOAMevol} we construct and solve the evolution equations for the first impact parameter moment of the amplitude. The consequences of this solution for quark OAM at small $x$ are summarized in Sec.~\ref{sec:quarkOAMres}, with the resulting small-$x$ asymptotics of quark OAM distribution given by \eq{e:MAINRESULT2} (at large $N_c$ and in DLA). 

The gluon OAM distribution is analyzed in Sec.~\ref{sec:gluonOAM} following the same general strategy. Using the gluon Wigner distribution the gluon OAM operator is constructed in Sec.~\ref{sec:oper}. It is simplified at small $x$ in Sec.~\ref{sec:eval}. The gluon OAM distribution, just like the gluon helicity, is related to a different polarized gluon dipole operator defined in \cite{Hatta:2016aoc,Kovchegov:2017lsr}. Similar to the quark OAM case, the gluon OAM distribution is related to the first impact parameter moment of the polarized gluon dipole amplitude (see \eq{G3Gi}). The evolution equation for this moment is derived and solved in Sec.~\ref{sec:solution}. Evolution equations for gluon helicity are a bit more involved than those for quark helicity \cite{Kovchegov:2017lsr}: the same applies to OAMs. Finally, the solution is employed in Sec.~\ref{sec:result} to derive the small-$x$ asymptotics \eqref{e:MAINRESULT} of the gluon OAM distribution (in the DLA limit and at large $N_c$). 

The results of our analysis are concisely summarized in the equations \eqref{OAM_abs} in the Abstract above as well as in Sec.~\ref{sec:sum}.

%
\section{Quark OAM} 
\label{sec:quarkOAM}
%

\subsection{The quark OAM operator}
\label{sec:quarkOAMoperator}

We start with a generic (quark or gluon) OAM written in terms of the Wigner distribution function $W (p,b)$,
\begin{align}\label{OAM0}
L_z = \int \frac{d^2 b_\perp d b^- \, d^2 k_\perp \, d k^+}{(2 \pi)^3} \, ({\un b} \times {\un k})_z \, W (k,b).
\end{align} 
Our notation for the light-cone components of the 4-vectors is $v^\pm = (v^0 \pm v^3)/\sqrt{2}$, while the transverse vectors are defined as ${\un v} = (v^1, v^2)$. We also denote $b = (b^-, {\un b})$ and $k= (k^+, {\un k})$.

To construct the quark OAM operator using \eq{OAM0} we need the quark Wigner function. We can extract the Wigner function from the unpolarized quark transverse-momentum dependent (TMD) distribution (in a longitudinally polarized proton) \cite{Mulders:1995dh}
\begin{align} \label{e:fdef}
f_{1}^q (x, k_T^2) = \frac{1}{(2\pi)^3} \, \int d^2 r \, dr^- \,
e^{i k \cdot r} \bra{P S_L} \bar\psi(0) \, \mathcal{U}[0,r] \, \frac{\gamma^+}{2} \,
\psi(r) \ket{P S_L}_{r^+ = 0} ,
\end{align}
with $k^+ = x P^+$ and $\mathcal{U}[0,r]$ a Wilson line staple connecting points $r$ and $0$. At small-$x$ and in the $A^- =0$ gauge we write this TMD (in the case of the semi-inclusive deep inelastic scattering (SIDIS) future-pointing staple) as \cite{Kovchegov:2018znm}
\begin{align}\label{TMD11}
f_{1}^q (x, k_T^2) & = \frac{2 P^+}{(2\pi)^3} \sum_X \: \int d^{2} \zeta \, d \zeta^- \, d^{2} \xi \, d \xi^-
\, e^{i k \cdot (\zeta - \xi)} 
\left\langle \bar\psi_\alpha (\xi) \, V_{\ul \xi} [\xi^-, \infty]  \ket{X} \left( \thalf \gamma^+ \right)_{\alpha \beta} \,
 \bra{X}  V_{\ul \zeta} [\infty , \zeta^-] \, \psi_\beta (\zeta) \right\rangle
\notag \\ & 
\equiv \frac{P^+}{(2 \pi)^3} \, \int d^2 \left( \thalf \zeta_\perp + \thalf \xi_\perp \right) \, d \left( \thalf \zeta^- + \thalf \xi^- \right) \, W^{q, SIDIS} \left( k, \thalf \zeta + \thalf \xi \right).
\end{align}
Here the large angle brackets denote the averaging in the target (proton) wave function, as is done in the saturation/color glass condensate (CGC) physics \cite{Gribov:1984tu,Iancu:2003xm,Weigert:2005us,Jalilian-Marian:2005jf,Gelis:2010nm,Albacete:2014fwa,KovchegovLevin}. The averaging is discussed in Appendix~\ref{app:CGCave} in more detail and is given by \eq{matrix_el3} there. We also use the following notation for the fundamental Wilson lines on the $x^-$ light cone,
\begin{align}
  V_{\un{x}} [b^-, a^-] = \mathcal{P} \exp \left[ i g
    \int\limits_{a^-}^{b^-} d x^- \, A^+ (x^+=0, x^-, {\un x})
  \right].
\end{align}

The second line in \eq{TMD11} is the definition of the quark SIDIS Wigner distribution, which we can use to obtain
\begin{align}
W^{q, SIDIS} (k, b) = 2 \sum_X \: \int d^{2} r \, d r^- \, 
\, e^{i k \cdot r} 
\left\langle \bar\psi_\alpha \left( b - \thalf r \right) \, V_{{\un b} - \thalf {\un r}} \left[ b^- - \thalf r^-, \infty \right]  \ket{X} \left( \thalf \gamma^+ \right)_{\alpha \beta} \right. \\ \times \left. \,
 \bra{X}  V_{{\un b} + \thalf {\un r}} \left[ \infty , b^- + \thalf r^- \right] \, \psi_\beta \left( b + \thalf r \right) \right\rangle . \notag
\end{align}
Inserting this into \eq{OAM0} yields
\begin{align}\label{qOAM123}
L_q (Q^2) = \frac{2 P^+}{(2\pi)^3} \,  \sum_X \: \int d^2 k_\perp \, d x \, d^{2} \zeta \, d \zeta^- \, d^{2} \xi \, d \xi^-
\, e^{i k \cdot (\zeta - \xi)} \left( \frac{{\un \zeta} + {\un \xi}}{2} \times {\un k}\right)
\left\langle \bar\psi_\alpha (\xi) \, V_{\ul \xi} [\xi^-, \infty]  \ket{X}  \, \left( \thalf \gamma^+ \right)_{\alpha \beta} \right. \\ \times \left.  \,
 \bra{X}  V_{\ul \zeta} [\infty , \zeta^-] \, \psi_\beta (\zeta) \right\rangle . \notag
\end{align}
(Note that the $x$-integral from here on is assumed to run from 0 to 1.) We are particularly interested in the quark OAM distribution, $L_q (x, Q^2) = dL_q (Q^2)/dx$, which can be easily read from \eq{qOAM123},
\begin{align}\label{qOAM124}
L_q (x, Q^2) = \frac{2 P^+}{(2\pi)^3} \,  \sum_X \: \int d^2 k_\perp \, d^{2} \zeta \, d \zeta^- \, d^{2} \xi \, d \xi^-
\, e^{i k \cdot (\zeta - \xi)} \left( \frac{{\un \zeta} + {\un \xi}}{2} \times {\un k}\right)
\left\langle \bar\psi_\alpha (\xi) \, V_{\ul \xi} [\xi^-, \infty]  \ket{X}  \, \left( \thalf \gamma^+ \right)_{\alpha \beta} \right. \\ \times \left.  \,
 \bra{X}  V_{\ul \zeta} [\infty , \zeta^-] \, \psi_\beta (\zeta) \right\rangle . \notag
\end{align}

\subsection{Evaluation of the quark OAM operator at small $x$}
\label{sec:quarkOAMeval}

We now need to simplify the expression \eqref{qOAM124} for the quark OAM distribution at small $x$. Following the technique described in \cite{Kovchegov:2018znm} (see Eq.~(10) there) we write
\begin{align}\label{qOAM1}
L_q (x, Q^2) = \frac{2 P^+}{(2\pi)^3} \, \sum_{\bar q} \, \int d^2 k_\perp \, d^{2} \zeta \, d^{2} \xi \, & \int\limits_{-\infty}^0 d \zeta^- \, \int\limits^{\infty}_0 d \xi^- \, e^{i k \cdot (\zeta - \xi)} \left( \frac{{\un \zeta} + {\un \xi}}{2} \times {\un k}\right) \,  \left( \thalf \gamma^+ \right)_{\alpha \beta}\notag \\ & \times
\left\langle \bar\psi_\alpha (\xi) \, \ket{\bar q} \, \bra{\bar q}
 V_{\ul \zeta} [\infty , -\infty] \, \psi_\beta (\zeta) \right\rangle + \cc .
\end{align}
For $\zeta^- <0$ and $\xi^- >0$ one can write down the quark propagator through the shock wave as \cite{Kovchegov:2018znm,Kovchegov:2017lsr}
\begin{align}\label{propagator2}
\contraction
{}
{\bar\psi^i_\alpha}
{(\eta) \:}
{\psi^j_\beta}
\bar\psi^i_\alpha (\xi) \: \psi^j_\beta (\zeta)
&= \int  d^2 w \, \frac{d^2 k_1 \, d k_1^-}{(2\pi)^3} \, \frac{d^2 k_2}{(2\pi)^2} \, e^{i \frac{{\un k}_1^2}{2 k_1^-} \zeta^- - i \frac{{\un k}_2^2}{2 k_1^-} \xi^- + i \ul{k}_1 \cdot (\ul{w} - \ul{\zeta}) + i \ul{k}_2 \cdot (\ul{\xi} - \ul{w})} \, \theta (k_1^-)
\notag \\ & \hspace{2cm} \times
\left\{ \left[ \frac{\slashed{k_1}}{2 k_1^-} \right]
\left[ \left( \hat{V}_{{\un w}}^\dagger \right)^{ji} \right]
\left[ \frac{ \slashed{k_2}}{2 k_1^-} \right] \right\}_{\beta \alpha} \Bigg|_{k_2^- = k_1^-, k_1^2 =0, k_2^2 =0} 
\end{align}
such that
\begin{align}\label{qOAM3}
L_q (x, Q^2) = \frac{2 P^+}{(2\pi)^3} \, &  \int d^2 k_\perp \, d^{2} \zeta \, d^{2} \xi \, \int\limits_{-\infty}^0 d \zeta^- \, \int\limits^{\infty}_0 d \xi^- \, e^{i k \cdot (\zeta - \xi)} \left( \frac{{\un \zeta} + {\un \xi}}{2} \times {\un k}\right) \notag \\ & \times \int  d^2 w \, \frac{d^2 k_1 \, d k_1^-}{(2\pi)^3} \, \frac{d^2 k_2}{(2\pi)^2}  \, e^{i \frac{{\un k}_1^2}{2 k_1^-} \zeta^- - i \frac{{\un k}_2^2}{2 k_1^-} \xi^- + i \ul{k}_1 \cdot (\ul{w} - \ul{\zeta}) + i \ul{k}_2 \cdot (\ul{\xi} - \ul{w})}
 \, \theta (k_1^-) \, \left( \thalf \gamma^+ \right)_{\alpha \beta} \notag \\ & \times \, \left\langle \mbox{T} \,
 V_{\ul \zeta}^{ij} [\infty , -\infty] \, \left\{ \left[ \frac{\slashed{k_1}}{2 k_1^-} \right]
\left[ \left( \hat{V}_{{\un w}}^\dagger \right)^{ji} \right]
\left[ \frac{ \slashed{k_2}}{2 k_1^-} \right] \right\}_{\beta \alpha} \Bigg|_{k_2^- = k_1^-, k_1^2 =0, k_2^2 =0}  \right\rangle  + \mbox{c.c.}.
\end{align}
As in  \cite{Kovchegov:2018znm}, we are using the time-ordering sign T to delineate the amplitude from the complex conjugate amplitude, with the latter containing the anti-time ordering sign $\bar{\mbox{T}}$. 
Integrating over $\zeta^-$ and $\xi^-$  and neglecting higher powers of $x$ yields
\begin{align}\label{qOAM4}
L_q (x, Q^2) = - \frac{2 P^+}{(2\pi)^3} \, &  \int d^2 k_\perp \, d^{2} \zeta \, d^{2} \xi \, e^{- i {\un k} \cdot ({\un \zeta} - {\un \xi})} \left( \frac{{\un \zeta} + {\un \xi}}{2} \times {\un k}\right) \notag \\ & \times \int  d^2 w \, \frac{d^2 k_1 \, d k_1^-}{(2\pi)^3} \, \frac{d^2 k_2}{(2\pi)^2}  \, e^{ i \ul{k}_1 \cdot (\ul{w} - \ul{\zeta}) + i \ul{k}_2 \cdot (\ul{\xi} - \ul{w})}
 \, \theta (k_1^-) \, \left( \thalf \gamma^+ \right)_{\alpha \beta} \notag \\ & \times \, \left\langle \mbox{T} \,
 V_{\ul \zeta}^{ij} [\infty , -\infty] \, \left\{ \frac{\slashed{k_1}}{{\un k}_1^2}
\, \left( \hat{V}_{{\un w}}^\dagger \right)^{ji} \,
\frac{\slashed{k_2}}{{\un k}_2^2}  \right\}_{\beta \alpha} \Bigg|_{k_2^- = k_1^-, k_1^2 =0, k_2^2 =0}  \right\rangle  + \mbox{c.c.}.
\end{align}

Employing polarization sums we write
\begin{align}\label{qOAM5}
L_q (x, Q^2) = - \frac{2 P^+}{(2\pi)^3} \, &  \int d^2 k_\perp \, d^{2} \zeta \, d^{2} \xi \, e^{- i {\un k} \cdot ({\un \zeta} - {\un \xi})} \left( \frac{{\un \zeta} + {\un \xi}}{2} \times {\un k}\right) \notag \\ & \times \int  d^2 w \, \frac{d^2 k_1 \, d k_1^-}{(2\pi)^3} \, \frac{d^2 k_2}{(2\pi)^2}  \, e^{ i \ul{k}_1 \cdot (\ul{w} - \ul{\zeta}) + i \ul{k}_2 \cdot (\ul{\xi} - \ul{w})}
 \, \theta (k_1^-) \, \sum_{\sigma_1, \sigma_2} {\bar v}_{\sigma_2} (k_2)  \, \thalf \gamma^+ v_{\sigma_1} (k_1) \notag \\ & \times \, \left\langle  \mbox{T} \,
 V_{\ul \zeta}^{ij} [\infty , -\infty] \, \frac{1}{{\un k}_1^2 \, {\un k}_2^2} \, {\bar v}_{\sigma_1} (k_1)
\, \left( \hat{V}_{{\un w}}^\dagger \right)^{ji} \, v_{\sigma_2} (k_2) \Bigg|_{k_2^- = k_1^-, k_1^2 =0, k_2^2 =0}  \right\rangle  + \mbox{c.c.}.
\end{align}

The polarized ``Wilson line" is defined by \cite{Kovchegov:2018znm}
\begin{align}\label{spinors1}
\left[ {\bar v}_{\sigma} (p) \, \Big( \hat{V}^\dagger_{\ul x} \Big) \, v_{\sigma'} (p') \right] &=
2 \sqrt{p^- p^{\prime \, -}} \delta_{\sigma \sigma'} \left( V^\dagger_{\ul x} - \sigma V_{\ul x}^{pol \, \dagger} + \ldots \right) 
= 2 \sqrt{p^- p^{\prime \, -}} \delta_{\sigma \sigma'} \: V^\dagger_{\ul x} (-\sigma) + \ldots ,
\end{align}
where ellipsis denote polarization-independent sub-eikonal terms, which are not important for our calculation. Here we employ an abbreviated notation $V_{\un x} \equiv V_{\un x} [\infty, -\infty]$. We will use the 
$+ \leftrightarrow -$ interchanged Brodsky-Lepage (BL) spinors \cite{Lepage:1980fj}, which we will also refer to as the anti-BL spinors \cite{Kovchegov:2018znm,Kovchegov:2018zeq}:
\begin{align}
u_\sigma (p) = \frac{1}{\sqrt{\sqrt{2} \, p^-}} \, [\sqrt{2} \, p^- + m \, \gamma^0 +  \gamma^0 \, {\un \gamma} \cdot {\un p} ] \,  \rho (\sigma), \ \ \ v_\sigma (p) = \frac{1}{\sqrt{\sqrt{2} \, p^-}} \, [\sqrt{2} \, p^- - m \, \gamma^0 +  \gamma^0 \, {\un \gamma} \cdot {\un p} ] \,  \rho (-\sigma),
\end{align}
where $p^\mu = \left( \frac{{\un p}^2+ m^2}{2 p^-}, p^-, {\un p} \right)$ and
\begin{align}
  \rho (+1) \, = \, \frac{1}{\sqrt{2}} \, \left(
  \begin{array}{c}
      1 \\ 0 \\ -1 \\ 0
  \end{array}
\right), \ \ \ \rho (-1) \, = \, \frac{1}{\sqrt{2}} \, \left(
  \begin{array}{c}
        0 \\ 1 \\ 0 \\ 1
  \end{array}
\right) .
\end{align}
Using these spinors with massless quarks we get
\begin{align}\label{spinors2}
{\bar v}_{\sigma_2} (k_2) \thalf \gamma^+ v_{\sigma_1} (k_1)  =  \thalf 
\delta_{\sigma_2 \sigma_1} 
\, \frac{(\ul{k}_2 \cdot \ul{k}_1) + i \sigma_1 (\ul{k}_1 \times \ul{k}_2)}
{\sqrt{k_1^- k_2^-}}  .
\end{align}
Performing the sum over $\sigma_1, \sigma_2$ in \eq{qOAM5} with the help of Eqs.~\eqref{spinors1} and \eqref{spinors2} we arrive at
\begin{align}\label{qOAM6}
L_q (x, Q^2) = - \frac{4 P^+}{(2\pi)^3} \, &  \int d^2 k_\perp \, d^{2} \zeta \, d^{2} \xi \, e^{- i {\un k} \cdot ({\un \zeta} - {\un \xi})} \left( \frac{{\un \zeta} + {\un \xi}}{2} \times {\un k}\right) \notag \\ & \times \int  d^2 w \, \frac{d^2 k_1 \, d k_1^-}{(2\pi)^3} \, \frac{d^2 k_2}{(2\pi)^2}  \, e^{ i \ul{k}_1 \cdot (\ul{w} - \ul{\zeta}) + i \ul{k}_2 \cdot (\ul{\xi} - \ul{w})}
 \, \theta (k_1^-) \, \frac{1}{{\un k}_1^2 \, {\un k}_2^2} \, \notag \\ & \times \, \left\langle {\un k}_1 \cdot {\un k}_2 \,
 \mbox{T} \, \tr \left[ V_{\ul \zeta} \, V^\dagger_{\ul w} \right]  - i {\un k}_1 \times {\un k}_2 \,
 \mbox{T} \, \tr \left[ V_{\ul \zeta} \, V^{pol \, \dagger}_{\ul w} \right] \right\rangle  + \mbox{c.c.}.
\end{align}
Next we integrate over ${\un k}_1$ and ${\un k}_2$. This yields
\begin{align}\label{qOAM7}
L_q (x, Q^2) = - \frac{4 P^+}{(2\pi)^5} \, &   \int d^2 k_\perp \, d^{2} \zeta \, d^{2} \xi \, d^2 w \, e^{- i {\un k} \cdot ({\un \zeta} - {\un \xi})} \left( \frac{{\un \zeta} + {\un \xi}}{2} \times {\un k}\right) \int\limits_0^\infty \frac{d k_1^-}{2\pi}  \notag \\ & \times  \, \left\{ \frac{\ul{\zeta} - {\un w}}{|\ul{\zeta} - {\un w}|^2} \cdot \frac{\ul{\xi} - {\un w}}{|\ul{\xi} - {\un w}|^2} \,
 \left\langle \mbox{T} \, \tr \left[ V_{\ul \zeta} \, V^\dagger_{\ul w} \right] \right\rangle  - i \frac{\ul{\zeta} - {\un w}}{|\ul{\zeta} - {\un w}|^2} \times \frac{\ul{\xi} - {\un w}}{|\ul{\xi} - {\un w}|^2} \,
\left\langle \mbox{T} \, \tr \left[ V_{\ul \zeta} \, V^{pol \, \dagger}_{\ul w} \right] \right\rangle \right\} + \mbox{c.c.}.
\end{align}
Adding the complex conjugate we obtain
\begin{align}\label{qOAM8}
& \, L_q (x, Q^2) = - \frac{4 P^+}{(2\pi)^5} \, \int d^2 k_\perp \, d^{2} \zeta \, d^{2} \xi \, d^2 w \, e^{- i {\un k} \cdot ({\un \zeta} - {\un \xi})} \left( \frac{{\un \zeta} + {\un \xi}}{2} \times {\un k}\right) \int\limits_0^\infty \frac{d k_1^-}{2\pi}   \\ & \times  \, \left\{ \frac{\ul{\zeta} - {\un w}}{|\ul{\zeta} - {\un w}|^2} \cdot \frac{\ul{\xi} - {\un w}}{|\ul{\xi} - {\un w}|^2} \,
 \left\langle \mbox{T} \, \tr \left[ V_{\ul \zeta} \, V^\dagger_{\ul w} \right] + \bar{\mbox{T}} \, \tr \left[ V_{\ul w} \, V^\dagger_{\ul \xi} \right] \right\rangle - i \frac{\ul{\zeta} - {\un w}}{|\ul{\zeta} - {\un w}|^2} \times \frac{\ul{\xi} - {\un w}}{|\ul{\xi} - {\un w}|^2} \,
\left\langle \mbox{T} \, \tr \left[ V_{\ul \zeta} \, V^{pol \, \dagger}_{\ul w} \right] + \bar{\mbox{T}} \, \tr \left[  V^{pol}_{\ul w} \, V_{\ul \xi}^\dagger \right] \right\rangle \right\} . \notag
\end{align}
In the second term of each angle brackets we replace ${\un k} \to - {\un k}$ and interchange ${\un \zeta} \leftrightarrow {\un \xi}$ :
\begin{align}\label{qOAM9}
& \, L_q (x, Q^2) = - \frac{4 P^+}{(2\pi)^5} \, \int d^2 k_\perp \, d^{2} \zeta \, d^{2} \xi \, d^2 w \, e^{- i {\un k} \cdot ({\un \zeta} - {\un \xi})} \left( \frac{{\un \zeta} + {\un \xi}}{2} \times {\un k}\right) \int\limits_0^\infty \frac{d k_1^-}{2\pi}   \\ & \times  \, \left\{ \frac{\ul{\zeta} - {\un w}}{|\ul{\zeta} - {\un w}|^2} \cdot \frac{\ul{\xi} - {\un w}}{|\ul{\xi} - {\un w}|^2} \,
 \left\langle \mbox{T} \, \tr \left[ V_{\ul \zeta} \, V^\dagger_{\ul w} \right] - \bar{\mbox{T}} \, \tr \left[ V_{\ul w} \, V^\dagger_{\ul \zeta} \right] \right\rangle  -  i \frac{\ul{\zeta} - {\un w}}{|\ul{\zeta} - {\un w}|^2} \times \frac{\ul{\xi} - {\un w}}{|\ul{\xi} - {\un w}|^2} \,
\left\langle \mbox{T} \, \tr \left[ V_{\ul \zeta} \, V^{pol \, \dagger}_{\ul w} \right] + \bar{\mbox{T}} \, \tr \left[  V^{pol}_{\ul w} \, V_{\ul \zeta}^\dagger \right] \right\rangle \right\} . \notag
\end{align}
Employing the reflection symmetry with respect to the final-state cut, or, equivalently using Eqs.~(22) from \cite{Kovchegov:2018znm} we conclude that (cf. Eq.~(24) in \cite{Kovchegov:2018znm})
\begin{align}
\left\langle \mbox{T} \, \tr \left[ V_{\ul \zeta} \, V^\dagger_{\ul w} \right] - \bar{\mbox{T}} \, \tr \left[ V_{\ul w} \, V^\dagger_{\ul \zeta} \right] \right\rangle = \left\langle  \tr \left[ V_{\ul \zeta} \, V^\dagger_{\ul w} \right] - \tr \left[  V^\dagger_{\ul \zeta} \, V_{\ul w} \right] \right\rangle = 0,
\end{align}
where the last step employed the same reflection symmetry, which has been verified up to NLO in the unpolarized small-$x$ evolution \cite{Mueller:2012bn}. We are thus left with
\begin{align}\label{qOAM9b}
& \, L_q  (x, Q^2) = \frac{4 P^+ i}{(2\pi)^5} \, \int d^2 k_\perp \, d^{2} \zeta \, d^{2} \xi \, d^2 w \, e^{- i {\un k} \cdot ({\un \zeta} - {\un \xi})} \left( \frac{{\un \zeta} + {\un \xi}}{2} \times {\un k}\right) \int\limits_0^\infty \frac{d k_1^-}{2\pi}   \\ & \times  \, \left[ \frac{\ul{\zeta} - {\un w}}{|\ul{\zeta} - {\un w}|^2} \times \frac{\ul{\xi} - {\un w}}{|\ul{\xi} - {\un w}|^2}  \right] \,
\left\langle \mbox{T} \, \tr \left[ V_{\ul \zeta} \, V^{pol \, \dagger}_{\ul w} \right] + \bar{\mbox{T}} \, \tr \left[  V^{pol}_{\ul w} \, V_{\ul \zeta}^\dagger \right] \right\rangle . \notag
\end{align}
For the flavor-singlet case we need to add the antiquark contribution. This yields 
\begin{align}\label{qOAM9c}
& \, L_{q + \bar{q}}  (x, Q^2) = \frac{4 P^+ i}{(2\pi)^5} \, \int d^2 k_\perp \, d^{2} \zeta \, d^{2} \xi \, d^2 w \, e^{- i {\un k} \cdot ({\un \zeta} - {\un \xi})} \left( \frac{{\un \zeta} + {\un \xi}}{2} \times {\un k}\right) \int\limits_0^\infty \frac{d k_1^-}{2\pi}   \\ & \times  \, \left[ \frac{\ul{\zeta} - {\un w}}{|\ul{\zeta} - {\un w}|^2} \times \frac{\ul{\xi} - {\un w}}{|\ul{\xi} - {\un w}|^2} \right] \,
\left\langle \mbox{T} \, \tr \left[ V_{\ul \zeta} \, V^{pol \, \dagger}_{\ul w} \right] + \mbox{T} \, \tr \left[ V^{pol}_{\ul w} \, V_{\ul \zeta}^\dagger \right] + \bar{\mbox{T}} \, \tr \left[  V^{pol}_{\ul w} \, V_{\ul \zeta}^\dagger \right] + \bar{\mbox{T}} \, \tr \left[  V_{\ul \zeta} V^{pol \, \dagger}_{\ul w}   \right] \right\rangle . \notag
\end{align}

Using the definition of the polarized dipole amplitude \cite{Kovchegov:2018znm}
\begin{align}
G_{{\un w}, {\un \zeta}} (zs) = \frac{k_1^- \, P^+}{N_c} \, \mbox{Re} \, \left\langle \mbox{T} \,  \mbox{tr} \left[ V_{\ul \zeta} \, V_{{\un w}}^{\dagger \, pol} \right] + \mbox{T} \,  \mbox{tr} \left[ V_{{\un w}}^{pol} \, V_{\ul \zeta}^\dagger \right] \right\rangle
\end{align}
with $z s = 2 k_1^- \, P^+$ and inserting proper limits of the $k_1^-$ integration we rewrite the contribution of \eq{qOAM9c} as (see \cite{Kovchegov:2018znm} for details)
\begin{align}\label{qOAM10}
L_{q + \bar{q}} (x, Q^2)  =& \, \frac{8 i N_c}{(2\pi)^6} \, \int d^2 k_\perp \, d^{2} \zeta \, d^{2} \xi \, d^2 w \, e^{- i {\un k} \cdot ({\un \zeta} - {\un \xi})} \left( \frac{{\un \zeta} + {\un \xi}}{2} \times {\un k}\right) \int\limits_{\Lambda^2/s}^{Q^2/(xs) \approx 1} \frac{d z}{z}  \,  \frac{\ul{\zeta} - {\un w}}{|\ul{\zeta} - {\un w}|^2} \times \frac{\ul{\xi} - {\un w}}{|\ul{\xi} - {\un w}|^2} \,
G_{{\un w}, {\un \zeta}} (zs) .
\end{align}
Here $s \approx Q^2/x$ is the center-of-mass energy squared for the polarized dipole--target system, $z$ is the minus momentum fraction of the dipole momentum carried by the soft quark or anti-quark line, and $\Lambda$ is the infrared (IR) cutoff. 

The expression \eqref{qOAM10} can be integrated over $\un \xi$:
\begin{align}\label{qOAM11}
L_{q + \bar{q}} (x, Q^2) =& \, \frac{8 i N_c}{(2\pi)^6} \int d^2 k_\perp \, d^{2} \zeta \, d^2 w \, e^{- i {\un k} \cdot ({\un \zeta} - {\un w})} \!  \int\limits_{\Lambda^2/s}^1 \frac{d z}{z}  \left[ 2 \pi i \, \frac{\ul{\zeta} - {\un w}}{|\ul{\zeta} - {\un w}|^2} \times \frac{{\un k}}{{\un k}^2} \, \left( \frac{{\un \zeta} + {\un w}}{2} \times {\un k} \right) - \pi \, \frac{\ul{\zeta} - {\un w}}{|\ul{\zeta} - {\un w}|^2} \cdot \frac{{\un k}}{{\un k}^2} \right] G_{{\un w}, {\un \zeta}} (zs) .
\end{align}

If we replace ${\un w} \to {\un x}_1$ and ${\un \zeta} \to {\un x}_0$, and use the integration variables ${\un x}_{10} = {\un x}_1 - {\un x}_0$ and ${\un x}_1$, then \eq{qOAM11} can be rewritten as
\begin{align}\label{qOAM14.5}
L_{q + \bar{q}} (x, Q^2)  =& \,  \frac{8 N_c}{(2\pi)^5} \, \int  \, d^2 k_\perp \, d^{2} x_{10} \, d^2 x_1 \, e^{i {\un k} \cdot {\un x}_{10}} \, \frac{{\un x}_{10}}{x_{10}^2} \times \frac{{\un k}}{{\un k}^2} \ {\un x}_1 \times {\un k} \ \int\limits_{\Lambda^2/s}^{1} \frac{d z}{z} \, G_{10} (zs) -  \sum_f [ \Delta q^f (x, Q^2)  +  \Delta \bar{q}^f (x, Q^2) ],
\end{align}
where we have summed over flavors and, for each flavor, 
\begin{align}
\Delta q^f (x, Q^2)  +  \Delta \bar{q}^f (x, Q^2) = \int d^2 k_\perp \, g_{1L}^S (x, k_T^2) 
\end{align}
with the flavor-singlet SIDIS quark helicity TMD at small $x$ \cite{Kovchegov:2018znm}
\begin{align}\label{TMD22}
g_{1L}^S (x, k_T^2) = 
 \frac{8 i N_c}{(2\pi)^5}  \: \int d^{2} \zeta 
\, d^2 w \, e^{ - i {\un k} \cdot (\ul{\zeta} - \ul{w})} \, \int\limits_{\Lambda^2/s}^1 \frac{d z}{z}  \,\frac{\un{\zeta} - \un{w}}{|\un{\zeta} - \un{w}|^2} \cdot  \frac{\un{k}}{{\un k}^2} \  G_{{\un w}, {\un \zeta}} (zs) .
\end{align}
In arriving at \eqref{qOAM14.5} we have used the fact that, for fixed ${\un x}_{10}$, the $x_1$-integral
\begin{align}\label{Gintb}
\int d^2 x_1 \, G_{10} (zs) \equiv G (x_{10}^2, zs)
\end{align}
is a function of $x_{10}^2$ only. (In our notation ${\un x}_{ij} = {\un x}_i - {\un x}_j$ and $x_{ij} = |{\un x}_i - {\un x}_j|$ for any $i,j$.)

At this point it may be tempting to conclude that since the small-$x$/large-$z s$ asymptotics of quark helicity distribution $\Delta q^f (x, Q^2)$ and $G (x_{10}^2, zs)$ were derived in \cite{Kovchegov:2016weo,Kovchegov:2017jxc}, then the small-$x$ asymptotics of the quark OAM distribution $L_{q + \bar{q}} (x, Q^2)$ follows straightforwardly from \eq{qOAM14.5}. This is almost correct, with one caveat: in \cite{Kovchegov:2016weo,Kovchegov:2017jxc} we found the asymptotics of $G (x_{10}^2, zs)$, as defined in \eq{Gintb}, that is, of $G_{10} (zs)$ integrated over all impact parameters, since this is what $\Delta \Sigma (x, Q^2)$ depends on. In \eq{qOAM14.5}, in the first term on its right-hand side, we need a different object: we need the ``first moment" of $G_{10} (zs)$ in the impact parameter (${\un x}_1$) space,
\begin{align}\label{moment1}
\int d^2 x_1 \, x_1^k \, G_{10} (z s) . 
\end{align} 
Here the index $k=1,2$. Our next step is to determine the small-$x$/large-$z s$  asymptotics of the ``moment" in \eq{moment1}.

\subsection{Evolution equations for quark OAM and their solution}
\label{sec:quarkOAMevol}

Define
\begin{subequations}\label{quarkOAM_moments}
\begin{align}
& I^k ({\un x}_{10}, z s)  = \int d^2 x_1 \, x_1^k \, G_{10} (z s), \label{Ikdef} \\ 
& J^k ({\un x}_{10}, x_{21}^2, z s)  =  \int d^2 x_1 \, x_1^k \, \Gamma_{10,21} (z s).
\end{align}
\end{subequations}
The evolution for these new objects in the large-$N_c$ DLA approximation can be found from Eqs.~(80) and (82) of \cite{Kovchegov:2015pbl} for the polarized dipole amplitude $G_{10} (z)$ and an auxiliary function, the polarized neighbor dipole amplitude $\Gamma_{10,21} (z')$ \cite{Kovchegov:2015pbl,Kovchegov:2018znm} (with the $S$-matrix for the unpolarized dipole amplitude taken to be $S=1$ in those equations):
\begin{subequations}\label{GGamma_eqs}
\begin{align}\label{GNc1}
G_{10} (z s) = G_{10}^{(0)} (z s) + \frac{\alpha_s \, N_c}{2 \pi^2} & \int\limits_{\frac{1}{s \, x_{10}^2}}^{z}
\frac{dz'}{z'} \int \frac{d^2 x_{2}}{x_{21}^2} \: \theta (x_{10} - x_{21}) \, \theta \left(x_{21}^2 - \frac{1}{z' s} \right)  \,   \left[ \Gamma_{10,21} (z' s) + 3 \, G_{21} (z' s) \right], \\
\Gamma_{10,21} (z' s) = G_{10}^{(0)} (z' s) + \frac{\alpha_s \, N_c}{2 \pi^2} & \int\limits_{\min \{ \Lambda^2, \frac{1}{x_{10}^2} \} / s }^{z'}
\frac{dz''}{z''} \int \frac{d^2 x_{3}}{x_{32}^2} \:   \theta \left( \min \{ x_{10}^2, x_{21}^2 z'/z'' \} - x_{32}^2 \right) \, \theta \left(x_{32}^2 - \frac{1}{z'' s} \right) \notag \\ & \times \,   \left[  \Gamma_{10,32} (z'' s) + 3 \, G_{32} (z'' s)  \right] .
\end{align}
\end{subequations}
Multiplying both sides by $x_1^k$ and integrating over $x_1$ while keeping ${\un x}_{10}$ fixed we get 
\begin{subequations}\label{Ievol2}
\begin{align}
& I^k ({\un x}_{10}, z s)  = I^{(0) \, k} ({\un x}_{10}, z s) + \frac{\alpha_s \, N_c}{2 \pi^2}  \int\limits_{\frac{1}{s \, x_{10}^2}}^{z}
\frac{dz'}{z'} \int \frac{d^2 x_{21}}{x_{21}^2} \: \theta (x_{10} - x_{21}) \, \theta \left(x_{21}^2 - \frac{1}{z' s} \right)  \, J^k ({\un x}_{10}, x_{21}^2, z' s) ,  \\ 
& J^k ({\un x}_{10}, x_{21}^2, z' s)  = {I}^{(0) \, k} ({\un x}_{10}, z' s) + \frac{\alpha_s \, N_c}{2 \pi^2}  \int\limits_{\min \{ \Lambda^2, \frac{1}{x_{10}^2} \} / s }^{z'}
\frac{dz''}{z''} \int \frac{d^2 x_{32}}{x_{32}^2} \:   \theta \left( \min \{ x_{10}^2, x_{21}^2 z'/z'' \} - x_{32}^2 \right) \, \theta \left(x_{32}^2 - \frac{1}{z'' s} \right)  \notag \\ & \hspace*{9cm} \times \, J^k ({\un x}_{10}, x_{32}^2, z'' s) .
\end{align}
\end{subequations}
In arriving at Eqs.~\eqref{Ievol2} we have neglected terms like $I^k ({\un x}_{21}, z' s)$, which are zero after the angular integration over the directions of ${\un x}_{21}$.

The inhomogeneous terms in Eqs.~\eqref{Ievol2} are
\begin{align}\label{initI}
I^{(0) \, k} ({\un x}_{10}, z s)  = \int d^2 x_1 \, x_1^k \, G_{10}^{(0)} (z s) ,
\end{align}
where, again, the integration is performed with fixed ${\un x}_{10}$. The Born-level initial conditions for the polarized dipole amplitude are (see Eq.~(13a) in \cite{Kovchegov:2016zex}, which assumes the polarized target to be a single quark at the origin ${\un 0}$ in the transverse plane) 
\begin{align}\label{initG}
G_{10}^{(0)} (z s)= \frac{\as^2 \, C_F}{2 N_c} \, \left[ \frac{C_F}{x_1^2} - 2 \pi \delta^2 ({\un x}_{1}) \, \ln (z s x_{10}^2) \right].
\end{align} 
Using \eq{initG} in \eq{initI} while assuming that the ${\un x}_1$-integral is regulated in the IR by an upper cutoff on the magnitude of ${\un x}_1$ yields
\begin{align}\label{initI1}
I^{(0) \, k} ({\un x}_{10}, z s) = \int d^2 x_1 \, x_1^k \, \theta \left( \frac{1}{\Lambda} - x_1 \right) \, \frac{\as^2 \, C_F}{2 N_c} \, \left[ \frac{C_F}{x_1^2} - 2 \pi \delta^2 ({\un x}_{1}) \, \ln (z s x_{10}^2) \right] =0.
\end{align}
With the zero inhomogeneous terms, Eqs.~\eqref{Ievol2} have a trivial solution:
\begin{align}\label{0sol}
I^k ({\un x}_{10}, z s) = 0, \ \ \ J^k ({\un x}_{10}, x_{21}^2, z s)  = 0. 
\end{align}
However, this conclusion changes for a slight variation of the IR regularization in \eq{initI1}. For instance, using $\theta \left[ \frac{1}{\Lambda} - ({\un x}_1 + {\un x}_0)/2 \right]$ gives a non-zero result,
\begin{align}\label{initI2}
I^{(0) \, k} ({\un x}_{10}, z s) = \int d^2 x_1 \, x_1^k \, \theta \left( \frac{1}{\Lambda} - \frac{{\un x}_1 + {\un x}_0}{2}  \right) \, \frac{\as^2 \, C_F}{2 N_c} \, \left[ \frac{C_F}{x_1^2} - 2 \pi \delta^2 ({\un x}_{1}) \, \ln (z s x_{10}^2) \right] = \frac{\as^2 \, \pi \, C_F^2}{4 N_c} \, x_{10}^k.
\end{align}
Therefore, we will proceed assuming that the inhomogeneous term $I^{(0) \, k} ({\un x}_{10}, z s)$ is not zero. As we will shortly see, the leading small-$x$ asymptotics of $L_{q + \bar{q}} (x, Q^2)$ is independent of whether $I^{(0) \, k} ({\un x}_{10}, z s)$ is zero or not. 

Using the fact that neither the initial condition \eqref{initI} nor the evolution equations \eqref{GGamma_eqs} contain a two-dimensional Levi-Civita symbol $\epsilon^{ij}$, we can write, without any loss of generality,
\begin{subequations}\label{ansatze1}
\begin{align}
& I^k ({\un x}_{10}, z s) = {x}_{10}^k \, I (x_{10}^2, z s), \label{Idef} \\
& J^k ({\un x}_{10}, x_{21}^2, z s)  = {x}_{10}^k \, J (x_{10}^2, x_{21}^2, z s).
\end{align}
\end{subequations}
Substituting Eqs.~\eqref{ansatze1} into Eqs.~\eqref{Ievol2} yields
\begin{subequations}\label{Ievol3}
\begin{align}
& I (x^2_{10}, z s)  = I^{(0)} ({x}^2_{10}, z s) + \frac{\alpha_s N_c}{2\pi} \, \int\limits_{\frac{1}{x_{10}^2 s}}^z
\frac{dz'}{z'} \int\limits^{x_{10}^2}_{\frac{1}{z' s}} \frac{d x_{21}^2}{x_{21}^2} \, J ({x}^2_{10}, x_{21}^2, z' s) ,  \\ 
& J (x^2_{10}, x_{21}^2, z' s)  = {I}^{(0)} (x^2_{10}, z' s) + \frac{\alpha_s N_c}{2\pi} \, \int\limits_{\frac{1}{x_{10}^2 s}}^{z'}
\frac{dz''}{z''} \int\limits^{\min \{ x_{10}^2, x_{21}^2 (z'/z'') \} }_{\frac{1}{z'' s}} \frac{d x_{32}^2}{x_{32}^2} \, J ({x}^2_{10}, x_{32}^2, z'' s)  .
\end{align}
\end{subequations}
Inspired by \eq{initI2} and by the prior experience \cite{Kovchegov:2016weo,Kovchegov:2017jxc,Kovchegov:2017lsr}, which demonstrated independence of small-$x$ asymptotics on the inhomogeneous term for helicity distributions, let us assume that $I^{(0)} ({x}^2_{10}, z s) = I^{(0)} ({x}^2_{10})$. In this case, defining
\begin{align}
{\bar I} (x^2_{10}, z s) = \frac{I (x^2_{10}, z s)}{I^{(0)} ({x}^2_{10})}, \ \ \ {\bar J} (x^2_{10}, x_{21}^2, z' s) = \frac{J (x^2_{10}, x_{21}^2, z' s)}{I^{(0)} ({x}^2_{10})},
\end{align}
we reduce Eqs.~\eqref{Ievol3} to 
\begin{subequations}\label{Ievol4}
\begin{align}
& {\bar I} (x^2_{10}, z s)  = 1 + \frac{\alpha_s N_c}{2\pi} \, \int\limits_{\frac{1}{x_{10}^2 s}}^z
\frac{dz'}{z'} \int\limits^{x_{10}^2}_{\frac{1}{z' s}} \frac{d x_{21}^2}{x_{21}^2} \, {\bar J} ({x}^2_{10}, x_{21}^2, z' s) ,  \\ 
& {\bar J} (x^2_{10}, x_{21}^2, z' s)  = 1 + \frac{\alpha_s N_c}{2\pi} \, \int\limits_{\frac{1}{x_{10}^2 s}}^{z'}
\frac{dz''}{z''} \int\limits^{\min \{ x_{10}^2, x_{21}^2 (z'/z'') \} }_{\frac{1}{z'' s}} \frac{d x_{32}^2}{x_{32}^2} \, {\bar J} ({x}^2_{10}, x_{32}^2, z'' s)  .
\end{align}
\end{subequations}

These equations are solved in Appendix~\ref{app:Born} (with ${\bar I} = {\bar G}_5$ and ${\bar J} = {\bar \Gamma}_5$ there, and with the $\beta=+1$ case of the solution in Appendix~\ref{app:Born} being of interest to us here). The resulting leading high-energy contribution is (cf. \eq{Isol10})
\begin{align}
{\bar I} (x^2_{10}, z)  = \frac{I_1 \left( 2 \sqrt{\frac{\alpha_s N_c}{2\pi}} \, \ln (z s x_{10}^2) \right)}{\sqrt{\frac{\alpha_s N_c}{2\pi}} \, \ln (z s x_{10}^2)} .
\end{align} 
We conclude that, for $z s x_{10}^2 \gg 1$, 
\begin{align}\label{Isol_full}
I (x^2_{10}, z)  = I^{(0)} ({x}^2_{10}, z) \, \frac{I_1 \left( 2 \sqrt{\frac{\alpha_s N_c}{2\pi}} \, \ln (z s x_{10}^2) \right)}{\sqrt{\frac{\alpha_s N_c}{2\pi}} \, \ln (z s x_{10}^2)} \sim  \left( z s x_{10}^2 \right)^{2 \sqrt{\frac{\alpha_s N_c}{2\pi}}} .
\end{align}


\subsection{Quark OAM distribution at small $x$}
\label{sec:quarkOAMres}

Employing Eqs.~\eqref{Ikdef} and \eqref{Idef} in \eq{qOAM14.5} we rewrite the quark OAM distribution as
\begin{align}\label{qOAM15}
L_{q + \bar{q}} (x, Q^2)  =& \,  \frac{8 N_c}{(2\pi)^5} \, \int  \, d^2 k_\perp \, d^{2} x_{10} \, e^{i {\un k} \cdot {\un x}_{10}} \, \frac{({\un x}_{10} \times {\un k})^2}{x_{10}^2 \, {\un k}^2} \, \int\limits_{\Lambda^2/s}^{1} \frac{d z}{z} \, I (x^2_{10}, zs) -  \sum_f [ \Delta q^f (x, Q^2)  +  \Delta \bar{q}^f (x, Q^2) ].
\end{align}
Equation~\eqref{Isol_full} allows us to conclude that the first term on the right-hand side of \eq{qOAM15} has the following small-$x$ asymptotics:
\begin{align}
\frac{8 N_c}{(2\pi)^5} \, \int  \, d^2 k_\perp \, d^{2} x_{10} \, e^{i {\un k} \cdot {\un x}_{10}} \, \frac{({\un x}_{10} \times {\un k})^2}{x_{10}^2 \, {\un k}^2} \, \int\limits_{\Lambda^2/s}^{1} \frac{d z}{z} \, I (x^2_{10}, zs) \sim \left( \frac{1}{x} \right)^{2 \sqrt{\frac{\alpha_s N_c}{2\pi}}} .
\end{align}
At the same time, the small-$x$ asymptotics of the quark helicity distribution was found in \cite{Kovchegov:2016weo,Kovchegov:2017jxc} to be
\begin{align}\label{q_hel}
\Delta \Sigma (x, Q^2) = \sum_f [ \Delta q^f (x, Q^2)  +  \Delta \bar{q}^f (x, Q^2) ] \sim \left(\frac{1}{x}\right)^{\alpha_h^q} =
  \left(\frac{1}{x}\right)^{\frac{4}{\sqrt{3}} \, \sqrt{\frac{\as
        \, N_c}{2 \pi}} } \approx \left(\frac{1}{x}\right)^{2.31 \,
    \sqrt{\frac{\as \, N_c}{2 \pi}}} 
\end{align}
in the same DLA limit. Since $4/\sqrt{3} >2$, we conclude that at small-$x$ the second term on the right-hand side of \eq{qOAM15} dominates. Dropping the first term we arrive at
\begin{align}\label{qOAM17}
L_{q + \bar{q}} (x, Q^2)  \approx - \sum_f \, \left[ \Delta q^f (x, Q^2)  +  \Delta \bar{q}^f (x, Q^2) \right] = - \Delta \Sigma (x, Q^2).
\end{align}
This result is in agreement with Eq.~(40) of \cite{Hatta:2018itc}, if we assume that $c = {\cal O} (\sqrt{\as}) \ll 1$ in it. Note, however, that the results in Sec.~IV of  \cite{Hatta:2018itc} (including Eq.~(40) there) were derived under the assumption that $|\Delta G | \gg |\Delta \Sigma |$ at small $x$, which is the opposite of what was found at DLA in \cite{Kovchegov:2016weo,Kovchegov:2017jxc,Kovchegov:2017lsr}.

The small-$x$ asymptotics of the quark OAM easily follows from Eqs.~\eqref{qOAM17} and \eqref{q_hel}. We conclude that
\begin{align} \label{e:MAINRESULT2}
 L_{q + \bar{q}} (x, Q^2)  = - \Delta \Sigma (x, Q^2) \sim 
  \left(\frac{1}{x}\right)^{\frac{4}{\sqrt{3}} \, \sqrt{\frac{\as
        \, N_c}{2 \pi}} } 
\end{align}
at small $x$ and in the large-$N_c$ limit (assuming gluon dominance in the latter). Note that the net small-$x$ quark contribution to the proton spin is
\begin{align}
\thalf \,\Delta \Sigma (x, Q^2) + L_{q + \bar{q}} (x, Q^2)  = - \thalf \,\Delta \Sigma (x, Q^2)
\end{align}
and is, therefore, non-zero.

%
\section{Gluon OAM} 
\label{sec:gluonOAM}
%

%
\subsection{The gluon OAM operator} \label{sec:oper}
%

Now we turn our attention to the gluon OAM distribution. First we need to construct the corresponding operator, and simplify it at small $x$. Again we begin with the definition of the OAM in terms of the Wigner function given in \eq{OAM0}. We need to obtain the gluon Wigner distribution. 

Similar to the quark case, to construct the gluon Wigner function we first consider the unpolarized dipole gluon TMD in a longitudinally polarized proton \cite{Dominguez:2011wm,KovchegovLevin}, 
\begin{align} 
\label{eq:TMDdef} 
f_{1}^{G \, dip} (x, k_T^2) = \frac{2}{x \, P^+} \, 
\, \int \frac{d \xi^- \, d^2\xi}{(2 \pi)^3} \, e^{i x P^+ \, \xi^- - i
  \un{k} \cdot \un{\xi}} \, \bra{P, S_L}  \tr \left[
  F^{+i} (0) \: {\cal U}^{[+]} [0,\xi] \: F^{+i} (\xi) \: {\cal
    U}^{[-]}[\xi, 0] \right] \ket{P, S_L}_{\xi^+ = 0},
\end{align}
where the future- and past-pointing Wilson line staples are ${\cal U}^{[+]} [0,\xi] = V_{\un 0} [0^-, +\infty] \, V_{\un \xi} [+\infty, \xi^-]$ and ${\cal
    U}^{[-]}[\xi, 0] = V_{\un \xi} [\xi^-, -\infty] \, V_{\un 0} [-\infty, 0^-]$ in $A^- =0$ gauge. 
To extract the gluon Wigner distribution we employ the CGC averaging in \eq{matrix_el1} to write
\begin{align}   
\label{eq:TMDdef2}
f_{1}^{G \, dip} (x, k_T^2) &= \frac{4}{x} \frac{1}{(2\pi)^3} \,
\int d \xi^- \, d^2\xi_\perp \: db^- \, d^2 b_\perp \:\: e^{i x P^+ \,
  \xi^-  - i \un{k} \cdot \un{\xi} }
\left\langle \tr \left[ F^{+i} (b) \: {\cal
      U}^{[+]}[b,b+\xi] \: F^{+i} (b+\xi) \: {\cal U}^{[-]} [b+\xi,
    b] \right] \right\rangle 
\notag \\ & 
\equiv \frac{P^+}{(2 \pi)^3} \, \int d^2 \left( b_\perp + \thalf \xi_\perp \right) \, d \left( b^- + \thalf \xi^- \right) \, W^{G \, dip} \left( k, b + \thalf \xi \right).
\end{align}
(The factor of $P^+$ is to ensure that the gluon PDF is per $dx$, not $d k^+$.) We read off the unpolarized gluon dipole Wigner distribution
\begin{align}\label{Wig3}
W^{G \, dip} (k,b) = \frac{4}{x P^+} \,  
& \int d \xi^- \, d^2\xi_\perp \:\: e^{i x P^+ \,
  \xi^-  - i \un{k} \cdot \un{\xi} } \\ \times & \,
\left\langle \tr \left[ F^{+i} (b- \thalf \xi) \: {\cal
      U}^{[+]}[b - \thalf \xi,b+ \thalf \xi] \: F^{+i} (b+\thalf \xi) \: {\cal U}^{[-]} [b+\thalf \xi,
    b - \thalf \xi] \right] \right\rangle. \notag
\end{align}
Using it in \eqref{OAM0} we arrive at the gluon dipole OAM definition
\begin{align}\label{OAM2}
L_G (Q^2) = \frac{4}{(2\pi)^3} \, & \int d^2 b_\perp d b^- \, d^2 k_\perp \, \frac{d x}{x} \, 
d \xi^- \, d^2\xi_\perp \ \left( {\un b} \times {\un k} \right) \  e^{i x P^+ \,
  \xi^-  - i \un{k} \cdot \un{\xi} }  \\ & \times \,
\left\langle \tr \left[ F^{+i} (b- \thalf \xi) \: {\cal
      U}^{[+]}[b- \thalf \xi,b+\thalf \xi] \: F^{+i} (b+\thalf \xi) \: {\cal U}^{[-]} [b+\thalf \xi,
    b - \thalf \xi] \right] \right\rangle . \notag
\end{align} 
In Appendix~\ref{sec:comparison} we show that this definition of gluon OAM is consistent with the standard Jaffe-Manohar gluon OAM definition \cite{Jaffe:1989jz}. 

Just like for quark OAM, we are interested in the gluon OAM distribution $L_G (x, Q^2) = d L_G (Q^2)/dx$, which is given by
\begin{align}\label{OAM2.5}
L_G (x, Q^2) = \frac{4}{(2\pi)^3 \, x} \, & \int d^2 b_\perp d b^- \, d^2 k_\perp \, 
d \xi^- \, d^2\xi_\perp \ \left( {\un b} \times {\un k} \right) \  e^{i x P^+ \,
  \xi^-  - i \un{k} \cdot \un{\xi} }  \\ & \times \,
\left\langle \tr \left[ F^{+i} (b- \thalf \xi) \: {\cal
      U}^{[+]}[b- \thalf \xi,b+\thalf \xi] \: F^{+i} (b+\thalf \xi) \: {\cal U}^{[-]} [b+\thalf \xi,
    b - \thalf \xi] \right] \right\rangle . \notag
\end{align} 
The presence of $\epsilon^{ij}$ in ${\un b} \times {\un k}$ of \eq{OAM2.5} demands that there has to be another $\epsilon^{ij}$ in the angle brackets $\langle \ldots \rangle$, thus eliminating the contributions of the standard (unpolarized) BFKL/BK/JIMWLK evolution. This is similar to the case of gluon helicity \cite{Kovchegov:2017lsr}.

%
\subsection{Evaluation of the gluon OAM operator at small $x$} \label{sec:eval}
%

Our next steps are to simplify the gluon OAM operator definition \eqref{OAM2} along the lines of \cite{Kovchegov:2017lsr,Hatta:2016aoc} and evolve it to small $x$.  In $A^- =0$ gauge \eq{OAM2.5} becomes
\begin{align}\label{OAM4}
L_G (x, Q^2) = \frac{4}{(2\pi)^3 \, x} \, & \int d \xi^- \, d^2\xi_\perp \ d \zeta^- d^2 \zeta_\perp \, d^2 k_\perp \, 
 \left( \frac{{\un \zeta} + {\un \xi}}{2} \times {\un k} \right) \  e^{i x P^+ \, (\xi^- - \zeta^-)
    - i \un{k} \cdot (\un{\xi} - \ul{\zeta})} \\ & \times \,
\left\langle \tr \left[ V_{\un \zeta} [-\infty, \zeta^-] \,  F^{+i} (\zeta) \, V_{\un \zeta} [ \zeta^-, +\infty] \, V_{{\un \xi}} [+\infty, \xi^-] \, F^{+i} (\xi) \,  V_{{\un \xi}} [\xi^-, -\infty]  \right] \right\rangle , \notag
\end{align} 
where we have also changed variables from $b \mp \thalf \xi \to \zeta, \xi$.

For the unpolarized gluon distribution, it is sufficient to replace the field-strength tensors by their eikonal approximations, $F^{+ i } \approx - \partial_\bot^i A^+$: however, in \eq{OAM4} this would give zero since the eikonal approximation contains no $\epsilon^{ij}$ needed to obtain a non-zero result. Hence we need to look for a sub-eikonal gluon field which (for mass-independent terms) depends on the polarization of the target proton, which would bring another $\epsilon^{ij}$. Proton polarization dependence enters through the sub-eikonal gluon field $A^i$ with $i=1,2$. The situation is similar to the case of gluon helicity at small $x$ \cite{Kovchegov:2017lsr}. We expand the
product of field-strength tensors to the first non-vanishing sub-eikonal order, that is, to the linear order in $A^i$, getting
\begin{align}
  & F^{+ i} (\zeta) \cdots F^{+ j} (\xi) = \! \Big( \partial^+ A_\bot^i (\zeta) - \partial^i A^+ (\zeta) - i g \, [A^+ (\zeta) \, , \, A_\bot^i (\zeta) ] \Big) \! \cdots \! \Big( \partial^+ A_\bot^j (\xi) - \partial^j A^+ (\xi) - i g \, [A^+ (\xi) \, , \, A_\bot^j (\xi) ] \Big) \\
  &\approx \left( \frac{\partial}{\partial \zeta^-} A_\bot^i (\zeta) -
    i g \, [A^+ (\zeta) \, , \, A_\bot^i (\zeta) ]\right) \cdots
  \left( \frac{\partial}{\partial \xi_\bot^j} A^+ (\xi) \right)
  + \left( \frac{\partial}{\partial \zeta_\bot^i} A^+ (\zeta) \right)
  \cdots \left( \frac{\partial}{\partial \xi^-} A_\bot^j (\xi) - i g
    \, [A^+ (\xi) \, , \, A_\bot^j (\xi) ]\right). \notag
\end{align}
We next convert the sub-eikonal part of the field-strength tensor
$F^{+ i} (\zeta)$ into a total derivative,
\begin{align}
  V_{\ul \zeta} [-\infty, \zeta^-] \left( \frac{\partial}{\partial
      \zeta^-} A_\bot^i (\zeta) - i g [ A^+ (\zeta) \, , \, A_\bot^i
    (\zeta) ] \right) V_{\ul \zeta} [\zeta^- , +\infty] =
  \frac{\partial}{\partial \zeta^-} \left( V_{\ul \zeta} [-\infty,
    \zeta^-] \: A_\bot^i (\zeta) \: V_{\ul \zeta} [\zeta^- , +\infty]
  \right) ,
\end{align}
which, after integration by parts, acts on the Fourier factor and
generates a net factor of $+ i x P^+$ on the right of \eq{OAM4}.  Analogously, the
sub-eikonal part of the $F^{+ j} (\xi)$ field-strength tensor gives
a net factor of $- i x P^+$ and the operator $A_\bot^j
(\xi)$.  After taking these derivatives, we set $e^{i x P^+
  (\xi^- - \zeta^-)} \approx 1$ in \eq{OAM4} (thus neglecting higher powers of
$x \ll 1$). We arrive at
\begin{align}\label{OAM5}
L_G (x, Q^2) = & \, \frac{4 i P^+}{(2\pi)^3} \, \int d \xi^- \, d^2\xi_\perp \ d \zeta^- d^2 \zeta_\perp \, d^2 k_\perp \,
 \left( \frac{{\un \zeta} + {\un \xi}}{2} \times {\un k} \right) \  e^{
    - i \un{k} \cdot (\un{\xi} - \ul{\zeta})} %
 \\ & \hspace{1cm} \times
\Bigg\{ \left\langle \tr \left[ V_{\ul \zeta} [-\infty, \zeta^-] \:
    A^i (\zeta) \: V_{\ul \zeta} [\zeta^-, +\infty] \:\: V_{\ul \xi}
    [+\infty, \xi^-] \: \left( \frac{\partial}{\partial \xi_\bot^i} A^+ (\xi) \right)
    \: V_{\ul \xi} [\xi^-, -\infty] \right] \right\rangle
\notag \\ & \hspace{1.5cm} -
\left\langle \tr \left[ V_{\ul \zeta} [-\infty, \zeta^-] \:
    \left( \frac{\partial}{\partial \zeta_\bot^i} A^+ (\zeta) \right) \: V_{\ul
      \zeta} [\zeta^-, +\infty] \:\: V_{\ul \xi} [+\infty, \xi^-] \:
    A^i (\xi) \: V_{\ul \xi} [\xi^-, -\infty] \right] \right\rangle
\Bigg\} . \notag
\end{align} 

Further, writing
\begin{align}
  \int\limits_{-\infty}^\infty d\zeta^- \, V_{\ul \zeta} [-\infty,
  \zeta^-] \, \left( \frac{\partial}{\partial \zeta_\bot^i} A^+ (\zeta) \right) \,
  V_{\ul \zeta} [\zeta^-, +\infty] = \frac{i}{g}
  \frac{\partial}{\partial \zeta_\bot^i} \, V_{\ul \zeta} [-\infty,
  +\infty] ,
\end{align}
yields
\begin{align}\label{OAM6}
L_G (x, Q^2) = & \, \frac{4 P^+}{g \, (2\pi)^3} \, \int d^2\xi_\perp \ d^2 \zeta_\perp \, d^2 k_\perp \,
 \left( \frac{{\un \zeta} + {\un \xi}}{2} \times {\un k} \right) \  e^{
    - i \un{k} \cdot (\un{\xi} - \ul{\zeta})} %
 \\ & \hspace{1cm} \times
\Bigg\{ \left\langle \tr \left[ \int d \zeta^- V_{\ul \zeta} [-\infty, \zeta^-] \:
    A^i (\zeta) \: V_{\ul \zeta} [\zeta^-, +\infty] \:\: \frac{\partial}{\partial \xi_\bot^i}  V_{\ul \xi}
    [+\infty, -\infty]  \right] \right\rangle
\notag \\ & \hspace{1.5cm} +
\left\langle \tr \left[ \frac{\partial}{\partial \zeta_\bot^i} \, V_{\ul \zeta} [-\infty,
  +\infty]  \:\: \int d \xi^- \, V_{\ul \xi} [+\infty, \xi^-] \:
    A^i (\xi) \: V_{\ul \xi} [\xi^-, -\infty] \right] \right\rangle
\Bigg\} . \notag
\end{align} 
Integrating by parts we obtain
\begin{align}\label{OAM7}
L_G (x, Q^2) = & \, \frac{4 P^+}{g \, (2\pi)^3} \, \int d^2\xi_\perp \ d^2 \zeta_\perp \, d^2 k_\perp \,
  e^{
    - i \un{k} \cdot (\un{\xi} - \ul{\zeta})} %
 \\ & \hspace{1cm} \times
\Bigg\{ i \, k^i \, \left( \frac{{\un \zeta} + {\un \xi}}{2} \times {\un k} \right) \ \left\langle \tr \left[ \int d \zeta^- V_{\ul \zeta} [-\infty, \zeta^-] \:
    A^i (\zeta) \: V_{\ul \zeta} [\zeta^-, +\infty] \:\:  V_{\ul \xi}
    [+\infty, -\infty]  \right] \right\rangle
\notag \\ & \hspace{1.5cm} - i \, k^i \, \left( \frac{{\un \zeta} + {\un \xi}}{2} \times {\un k} \right) \ 
\left\langle \tr \left[ V_{\ul \zeta} [-\infty,
  +\infty]  \:\:  \int d \xi^- \,  V_{\ul \xi} [+\infty, \xi^-] \:
    A^i (\xi) \: V_{\ul \xi} [\xi^-, -\infty] \right] \right\rangle
\notag \\ & \hspace{1.5cm} - \thalf \epsilon^{ij} \, k^j \, 
\left\langle \tr \left[ \int d \zeta^- V_{\ul \zeta} [-\infty, \zeta^-] \:
    A^i (\zeta) \: V_{\ul \zeta} [\zeta^-, +\infty] \:\:  V_{\ul \xi}
    [+\infty, -\infty]  \right] \right\rangle
\notag \\ & \hspace{1.5cm} - \thalf \epsilon^{ij} \, k^j \, 
\left\langle \tr \left[ V_{\ul \zeta} [-\infty,
  +\infty]  \:\:  \int d \xi^- \,  V_{\ul \xi} [+\infty, \xi^-] \:
    A^i (\xi) \: V_{\ul \xi} [\xi^-, -\infty] \right] \right\rangle
\Bigg\} . \notag
\end{align} 

Define the polarized Wilson line \cite{Hatta:2016aoc,Kovchegov:2017lsr}
\begin{align} \label{M:Vpol2} 
(V_{\ul x}^{pol})_\bot^i &\equiv
  \int\limits_{-\infty}^{+\infty} dx^- \, V_{\ul x} [+\infty, x^-] \:
  \left( i g \, P^+ \, A_\bot^i (x) \right) \: V_{\ul x} [x^- ,
  -\infty]
  \notag \\ &= \half \int\limits_{-\infty}^{+\infty} dx^- \, V_{\ul x}
  [+\infty, x^-] \: \left( i g \, \bar{A}_\bot^i (x) \right) \:
  V_{\ul x} [x^- , -\infty].
\end{align}
One may call it the polarized Wilson line of the second kind to stress its difference from a similar, but distinct, object defined for quark helicity and OAM (see also \cite{Kovchegov:2018znm}). With the help of \eq{M:Vpol2} we write
\begin{align}\label{OAM8}
L_G (x, Q^2) = & \ \frac{4}{g^2 \, (2\pi)^3} \,  \int d^2\xi_\perp \ d^2 \zeta_\perp \, d^2 k_\perp \, d x \ 
  e^{
    - i \un{k} \cdot (\un{\xi} - \ul{\zeta})} %
 \\ & \times
\Bigg\{ - k^i \, \left( \frac{{\un \zeta} + {\un \xi}}{2} \times {\un k} \right) \ \left\langle \tr \left[ (V_{\ul \zeta}^{pol \, \dagger})_\bot^i   \:\:  V_{\ul \xi}
    [+\infty, -\infty]  \right] \right\rangle
- k^i \, \left( \frac{{\un \zeta} + {\un \xi}}{2} \times {\un k} \right) \ 
\left\langle \tr \left[ V_{\ul \zeta} [-\infty,
  +\infty]  \:\: (V_{\ul \xi}^{pol})_\bot^i   \right] \right\rangle
\notag \\ & - \frac{i}{2} \,  \epsilon^{ij} \, k^j \, 
 \left\langle \tr \left[ (V_{\ul \zeta}^{pol \, \dagger})_\bot^i   \:\:  V_{\ul \xi}
    [+\infty, -\infty]  \right] \right\rangle 
+ \frac{i}{2} \,  \epsilon^{ij} \, k^j \, 
\left\langle \tr \left[ V_{\ul \zeta} [-\infty,
  +\infty]  \:\:  (V_{\ul \xi}^{pol})_\bot^i  \right] \right\rangle
\Bigg\} . \notag
\end{align} 
Swapping $\ul{\zeta} \leftrightarrow \ul{\xi}$ in the second and fourth terms in the curly brackets along with replacing ${\un k} \to - {\un k}$ for those terms we get
\begin{align}\label{OAM9}
L_G (x, Q^2) = & \ - \frac{4}{g^2 \, (2\pi)^3} \,  \int d^2\xi_\perp \ d^2 \zeta_\perp \, d^2 k_\perp \, d x \ 
  e^{
    - i \un{k} \cdot (\un{\xi} - \ul{\zeta})} %
 \\ & \times
\Bigg\{ k^i \, \left( \frac{{\un \zeta} + {\un \xi}}{2} \times {\un k} \right) \ \left\langle \tr \left[ (V_{\ul \zeta}^{pol \, \dagger})_\bot^i   \:\:  V_{\ul \xi}
    [+\infty, -\infty]  \right] \right\rangle
+ k^i \, \left( \frac{{\un \zeta} + {\un \xi}}{2} \times {\un k} \right) \ 
\left\langle \tr \left[ V_{\ul \xi} [-\infty,
  +\infty]  \:\: (V_{\ul \zeta}^{pol})_\bot^i   \right] \right\rangle
\notag \\ & + \frac{i}{2} \,  \epsilon^{ij} \, k^j \, 
 \left\langle \tr \left[ (V_{\ul \zeta}^{pol \, \dagger})_\bot^i   \:\:  V_{\ul \xi}
    [+\infty, -\infty]  \right] \right\rangle 
+ \frac{i}{2} \,  \epsilon^{ij} \, k^j \, 
\left\langle \tr \left[ V_{\ul \xi} [-\infty,
  +\infty]  \:\:  (V_{\ul \zeta}^{pol})_\bot^i  \right] \right\rangle
\Bigg\} . \notag
\end{align} 

Defining another polarized dipole-like operator \cite{Kovchegov:2017lsr}
\begin{align}   \label{eq:Gidef}
G^i_{10} (z s) \equiv \frac{1}{2 N_c} \,  
\left\langle \tr \left[ V_{\ul 0} (V_{\ul 1}^{pol \, \dagger})_\bot^i
\right] + \cc \right\rangle (z s)
\end{align}
and employing a more conventional (at small $x$) notation we rewrite \eq{OAM9} as
\begin{align}\label{OAM11}
L_G (x, Q^2) = & \ - \frac{8 N_c}{g^2 \, (2\pi)^3} \,  \int d^2 x_1 \ d^2 x_0 \, d^2 k_\perp \, 
  e^{i \un{k} \cdot {\un x}_{10}} 
\left[ k^i \, \left( \frac{{\un x}_1 + {\un x}_0}{2} \times {\un k} \right)  
+ \frac{i}{2} \,  \epsilon^{ij} \, k^j \right] \, G^i_{10} \left( z s = \frac{Q^2}{x} \right).
\end{align} 

Comparing this with the dipole gluon helicity TMD at small $x$ \cite{Kovchegov:2017lsr}
\begin{align}\label{GhTMD}
  g_{1L}^{G \, dip} (x, k_T^2) &= \frac{- 8 i \, N_c}{g^2 (2\pi)^3} \,
  \int d^2 x_1 \ d^2 x_0 \, e^{i \un{k} \cdot \un{x}_{10}} \: k_\bot^i
  \epsilon^{i j} \: G_{10}^j (z s =
    \tfrac{Q^2}{x}) 
\end{align}
we recast \eq{OAM11} as
\begin{align}\label{OAM13}
L_G (x, Q^2) = & \ - \frac{8 N_c}{g^2 \, (2\pi)^3} \,  \int d^2 x_1 \ d^2 x_0 \,  d^2 k_\perp \, 
  e^{
    i \un{k} \cdot {\un x}_{10}} 
\, k^i \, \left( \frac{{\un x}_1 + {\un x}_0}{2} \times {\un k} \right)  
 \, G^i_{10} \left( z s = \frac{Q^2}{x} \right) - \frac{1}{2} \, \Delta G (x,Q^2),
\end{align} 
where
\begin{align}
\Delta G (x,Q^2) = \int d^2  k_\perp \, g_{1L}^{G \, dip} (x, k_T^2). 
\end{align}

Next, write $k^i = - i \nabla^i_{x_1}$ and integrate by parts. This yields
\begin{align}\label{OAM14}
L_G (x, Q^2)  = & \ - \frac{8 N_c}{g^2 \, (2\pi)^3} \,  \int d^2 x_1 \ d^2 x_0 \,  d^2 k_\perp \, 
  e^{
    i \un{k} \cdot {\un x}_{10}} 
\, \left[ \frac{i}{2} \, \epsilon^{ij} \, k^j \, G^i_{10} +
 \left( \frac{{\un x}_1 + {\un x}_0}{2} \times {\un k} \right)  
 \, i \, \nabla^i_{x_1} G^i_{10} \right] - \frac{1}{2} \, \Delta G (x,Q^2),
\end{align} 
where we suppress the argument of $G^i_{10}$ for brevity. Using \eq{GhTMD} again we arrive at
\begin{align}\label{OAM15}
L_G (x, Q^2)  = & \ - \frac{8 N_c}{g^2 \, (2\pi)^3} \,  \int d^2 x_1 \ d^2 x_0 \,  d^2 k_\perp \, 
  e^{
    i \un{k} \cdot {\un x}_{10}} \,  \left( \frac{{\un x}_1 + {\un x}_0}{2} \times {\un k} \right)  
 \, i \, \nabla^i_{x_1} G^i_{10}  -  \Delta G (x,Q^2).
\end{align} 

Taking the Born-level $G^i_{10}$ from Eq.~(92) of \cite{Kovchegov:2017lsr} calculated for a single polarized quark target at $\un 0$,
\begin{align}\label{G10init}
  G^{i \, (0)}_{10} (z) = -
  \frac{\alpha_s^2 C_F}{N_c} \epsilon^{ij} \frac{({\un x}_1 - {\un
      b})^j}{|{\un x}_1 - {\un b}|^2} \, \ln \frac{|{\un x}_1 - {\un
      b}|}{|{\un x}_0 - {\un b}|},
\end{align} 
we get $\nabla^i_{x_1} G^i_{10} =0$. From \eq{OAM15} we see that at this Born level $L_G (x, Q^2) = - \Delta G (x,Q^2)$, in agreement with Eq.~(50) of \cite{Hatta:2016aoc} (after the latter is corrected by a factor of 2, as clarified in footnote 7 of \cite{Hatta:2018itc}).  This result appears to be similar to the parton model argument in Appendix~B of \cite{Hatta:2016aoc}. As we will see below, the Born level $L_G (x, Q^2) = - \Delta G (x,Q^2)$ relation does not appear to survive the DLA evolution. 

In the quark OAM case worked out above we learned that it is easier to work with the polarized dipole amplitude weighed by the position of the polarized quark ${\un x}_1$ and then integrated over all ${\un x}_1$, as opposed to using other weight factors (e.g. ${\un x}_0$ or $({\un x}_1 + {\un x}_0)/2$ as in \eq{OAM15}). To obtain the gluon OAM in terms of $x_1$-weighed polarized dipole amplitude, 
start with \eq{OAM13} and write $\frac{{\un x}_1 + {\un x}_0}{2} = {\un x}_1 - \thalf \, {\un x}_{10}$ along with replacing $d^2 x_1 \ d^2 x_0 \to d^2 x_1 \ d^2 x_{10}$. Then further replacing ${\un x}_{10} \to - i \, {\un \nabla}_k$ and integrating over $\un k$ by parts one arrives at
\begin{align}\label{OAM18}
L_G (x, Q^2) = - \frac{8 N_c}{g^2 \, (2\pi)^3} \,  \int d^2 x_1 \ d^2 x_{10} \,  d^2 k_\perp \, 
  e^{
    i \un{k} \cdot {\un x}_{10}} \, k^i \, \left( {\un x}_1 \times {\un k} \right)  
 \, G^i_{10} \left( z s = \frac{Q^2}{x} \right).
\end{align} 
This appears to be the most compact expression for the gluon OAM at small $x$. It also suggests that in the polarized dipole $01$ the two transverse coordinates do not enter on equal footing: this is indeed natural, since in \eq{eq:Gidef} line 1 is polarized, while line 0 is not. 

Further, we replace $k^i \to - i \nabla^i_{10}$, and, integrating by parts obtain
\begin{align}\label{OAM19}
L_G (x, Q^2) = & \ - \frac{8 i N_c}{g^2 \, (2\pi)^3} \,  \int d^2 x_1 \ d^2 x_{10} \,  d^2 k_\perp \, 
  e^{i \un{k} \cdot {\un x}_{10}} \, \left( {\un x}_1 \times {\un k} \right)  
 \, \nabla^i_{10} \, G^i_{10} \left( z s = \frac{Q^2}{x} \right).
\end{align} 

Consider a general decomposition 
\begin{align}
\int d^2 x_1 \ x_1^j \, \nabla^i_{10} \, G^i_{10} (zs) = x_{10}^j \, G_4 (x_{10}^2, zs) + \epsilon^{jk} x_{10}^k \, G_5 (x_{10}^2, zs)  . 
\end{align}
Note that the $x_1$ integration should be understood as keeping ${\un x}_{10}$ fixed, that is, ${\un x}_0 = {\un x}_1 - {\un x}_{10}$. 
Since $G^i_{10}$ contains exactly one $\epsilon^{ij}$ (see its evolution equations \eqref{M:evol4} below along with the initial conditions \eqref{G10init} or \eqref{CGC_Gi}), we conclude that $G_4 =0$ in the DLA and, therefore,
\begin{align}
\int d^2 x_1 \ x_1^j \, \nabla^i_{10} \, G^i_{10} (zs) = \epsilon^{jk} x_{10}^k \, G_5 (x_{10}^2 , zs) 
\end{align}
or, equivalently, 
\begin{align}\label{G3Gi}
G_5 (x_{10}^2 , zs)  = \frac{\epsilon^{jk} x_{10}^k}{x_{10}^2}  \int d^2 x_1 \ x_1^j \, \nabla^i_{10} \, G^i_{10} (zs). 
\end{align}
The gluon OAM becomes 
\begin{align}\label{OAM20}
L_G (x, Q^2) = - \frac{8 i N_c}{g^2 \, (2\pi)^3} \,  \int d^2 x_{10} \,  d^2 k_\perp \, 
  e^{i \un{k} \cdot {\un x}_{10}} \, \left( \un{k} \cdot {\un x}_{10} \right)  
 \, G_{5} \left( x_{10}^2, z s = \frac{Q^2}{x} \right).
\end{align} 

For comparison, the dipole gluon helicity TMD is (see \eq{GhTMD})
\begin{align}\label{Ghel_TMD}
g_{1L}^{G \, dip} (x, k_T^2) = \frac{8 i N_c}{g^2 \, (2\pi)^3} \,  \int d^2 x_{10} \, 
  e^{i \un{k} \cdot {\un x}_{10}} \, \left( \un{k} \cdot {\un x}_{10} \right)  
 \, G_{2} \left( x_{10}^2, z s = \frac{Q^2}{x} \right),
\end{align}
where
\begin{align}
\int d^2 x_1 \ G^i_{10} = \epsilon^{ij} x_{10}^j \, G_2 (x_{10}^2) , 
\end{align}
such that 
\begin{align}
G_2 (x_{10}^2)  = \frac{\epsilon^{jk} x_{10}^k}{x_{10}^2}  \int d^2 x_1 \, G^j_{10}. 
\end{align}

Similar to the quark helicity case, while the evolution equations for $G_{1 0}^{i} (z s)$ were constructed in \cite{Kovchegov:2017lsr}, their solution was found only for the impact parameter-integrated quantity $G_2$. Hence no solution for $G_{1 0}^{i} (z s)$ exists which would allow us to simply use \eq{G3Gi} to find $G_5$. Instead we need a relation between $G_5$ and $G_2$. To obtain it we need to construct a DLA evolution equation for $G_5 (x_{10}^2 , zs)$ first. Our next step is to use the evolution equations for $G_{1 0}^{i} (z s)$ derived in \cite{Kovchegov:2017lsr} to obtain the evolution equations for $G_5 (x_{10}^2 , zs)$ using \eq{G3Gi}. Note that, as pointed out above, at Born level $L_G = - \Delta G$, and, hence, $G_5^{(0)} = G_2^{(0)}$, which can also be verified independently by an explicit calculation.

%
\subsection{Evolution equations for gluon OAM and their solution} \label{sec:solution}
%

Start with Eqs.~(96) of \cite{Kovchegov:2017lsr}, 
\begin{subequations} \label{M:evol4}
\begin{align} 
  \label{M:evol4_1}
  G_{1 0}^{i} (z s) &= G_{1 0}^{i \, (0)} (z s) + \frac{\alpha_s
    N_c}{2\pi^2} \int\limits_{\frac{\Lambda^2}{s}}^z \frac{dz'}{z'}
  \int d^2 x_2 \, \ln\frac{1}{x_{21}^2 \Lambda^2} \: \frac{ \epsilon^{i
      j} \, x_{21}^j }{ x_{21}^2 } \: \Big[ \Gamma_{20 \, , \,
    21}^{gen} (z' s) + G_{21} (z' s) \Big]
\notag \\ &
- \frac{\alpha_s N_c}{2\pi^2} \int\limits_{\frac{\Lambda^2}{s}}^z
\frac{dz'}{z'} \int d^2 x_2 \, \ln\frac{1}{x_{21}^2 \Lambda^2} \: \frac{
  \epsilon^{i j} \, x_{20}^j }{ x_{20}^2 } \: \Big[
\Gamma_{20 \, , \, 21}^{gen} (z' s) + \Gamma_{21 \, , \, 20}^{gen} (z'
s) \Big]
\notag \\ &
+ \frac{\alpha_s N_c}{2\pi^2} \int\limits_{\frac{1}{x_{10}^2 s}}^z
\frac{dz'}{z'} \int \frac{d^2 x_2}{x_{21}^2} \: \theta\Big(x_{10}^2 -
x_{21}^2\Big) \, \theta\Big(x_{21}^2 - \frac{1}{z' s}\Big) \, \Big[
G_{12}^i (z' s) - \Gamma_{10 \, , \, 21}^i (z' s) \Big] ,
\\ \notag \\ \label{M:evol4_2}
\Gamma^i_{10, \, 21} (z' s) &= G_{10}^{i \, (0)} (z' s) +
\frac{\alpha_s N_c}{2\pi^2} \int\limits_{\frac{\Lambda^2}{s}}^{z'}
\frac{dz''}{z''} \int d^2 x_3 \, \ln\frac{1}{x_{31}^2 \Lambda^2} \: \frac{
  \epsilon^{i j} \, x_{31}^j }{x_{31}^2} \: \Big[ \Gamma_{30
  \, , \, 31}^{gen} (z'' s) + G_{31} (z'' s) \Big]
\notag \\ &
- \frac{\alpha_s N_c}{2\pi^2} \int\limits_{\frac{\Lambda^2}{s}}^{z'}
\frac{dz''}{z''} \int d^2 x_3 \, \ln\frac{1}{x_{31}^2 \Lambda^2} \: \frac{
  \epsilon^{i j} \, x_{30}^j }{x_{30}^2} \: \Big[ \Gamma_{30
  \, , \, 31}^{gen} (z'' s) + \Gamma_{31 \, , \, 30}^{gen} (z''
s)\Big]
\notag \\ &
+ \frac{\alpha_s N_c}{2\pi^2} \int\limits_{\frac{1}{x_{10}^2 s}}^{z'}
\frac{dz''}{z''} \int \frac{d^2 x_3}{x_{31}^2} \: \theta\Big(
\min\left[ x_{10}^2 \, , \, x_{21}^2 \tfrac{z'}{z''} \right] -
x_{31}^2 \Big) \: \theta\Big(x_{31}^2 - \frac{1}{z'' s} \Big) \, \Big[
G_{13}^i (z'' s) - \Gamma^i_{10 \, , \, 31} (z'' s) \Big] ,
\end{align}
\end{subequations}
with
\begin{align} 
  \label{Gen2}
  \Gamma^{gen}_{20,21} (z' s) = \theta (x_{20} - x_{21}) \,
  \Gamma_{20,21} (z' s) + \theta (x_{21} - x_{20}) \, G_{20} (z' s).
\end{align}

Employing \eq{G3Gi} we write for the first equation
\begin{align} 
  \label{G5evol}
  G_{5} (x_{10}^2, z s) &= G_{5}^{(0)} (x_{10}^2, z s) + \frac{\alpha_s
    N_c}{2\pi^2} \int\limits_{\frac{\Lambda^2}{s}}^z \frac{dz'}{z'} \, \frac{\epsilon^{km} x_{10}^m}{x_{10}^2} \nabla_{10}^i 
  \int d^2 x_2 \, d^2 x_1 \, x_1^k \, \ln\frac{1}{x_{21}^2 \Lambda^2} \: \frac{ \epsilon^{i
      j} \, x_{21}^j }{ x_{21}^2 } \: \Big[ \Gamma_{20 \, , \,
    21}^{gen} (z' s) + G_{21} (z' s) \Big]
\notag \\ &
- \frac{\alpha_s N_c}{2\pi^2} \int\limits_{\frac{\Lambda^2}{s}}^z
\frac{dz'}{z'} \, \frac{\epsilon^{km} x_{10}^m}{x_{10}^2} \nabla_{10}^i  \int d^2 x_2 \, d^2 x_1 \, x_1^k \, \ln\frac{1}{x_{21}^2 \Lambda^2} \: \frac{
  \epsilon^{i j} \, x_{20}^j }{ x_{20}^2 } \: \Big[
\Gamma_{20 \, , \, 21}^{gen} (z' s) + \Gamma_{21 \, , \, 20}^{gen} (z'
s) \Big]
\notag \\ &
+ \frac{\alpha_s N_c}{2\pi^2} \int\limits_{\frac{1}{x_{10}^2 s}}^z
\frac{dz'}{z'} \, \frac{\epsilon^{km} x_{10}^m}{x_{10}^2} \nabla_{10}^i \int \frac{d^2 x_2}{x_{21}^2} \: \theta\Big(x_{10}^2 -
x_{21}^2\Big) \, \theta\Big(x_{21}^2 - \frac{1}{z' s}\Big) \, d^2 x_1 \, x_1^k \, \Big[
G_{12}^i (z' s) - \Gamma_{10 \, , \, 21}^i (z' s) \Big] .
\end{align}
We stress that the $x_1$ integration should be understood as keeping ${\un x}_{10}$ fixed. In the last term in \eq{G5evol} we replace $d^2 x_2 \to d^2 x_{21}$. We also note that the operator $\nabla_{10}^i$ should not act on the first $\theta$-function, since this would lead to a non-DLA term. We thus arrive at 
\begin{align} 
  \label{G5evol2}
  G_{5} (x_{10}^2, z s) &= G_{5}^{(0)} (x_{10}^2, z s) + \frac{\alpha_s
    N_c}{2\pi^2} \int\limits_{\frac{\Lambda^2}{s}}^z \frac{dz'}{z'} \, \frac{\epsilon^{km} x_{10}^m}{x_{10}^2} \nabla_{10}^i 
  \int d^2 x_2 \, d^2 x_1 \, x_1^k \, \ln\frac{1}{x_{21}^2 \Lambda^2} \: \frac{ \epsilon^{i
      j} \, x_{21}^j }{ x_{21}^2 } \: \Big[ \Gamma_{20 \, , \,
    21}^{gen} (z' s) + G_{21} (z' s) \Big]
\notag \\ &
- \frac{\alpha_s N_c}{2\pi^2} \int\limits_{\frac{\Lambda^2}{s}}^z
\frac{dz'}{z'} \, \frac{\epsilon^{km} x_{10}^m}{x_{10}^2} \nabla_{10}^i  \int d^2 x_2 \, d^2 x_1 \, x_1^k \, \ln\frac{1}{x_{21}^2 \Lambda^2} \: \frac{
  \epsilon^{i j} \, x_{20}^j }{ x_{20}^2 } \: \Big[
\Gamma_{20 \, , \, 21}^{gen} (z' s) + \Gamma_{21 \, , \, 20}^{gen} (z'
s) \Big]
\notag \\ &
- \frac{\alpha_s N_c}{2\pi} \int\limits_{\frac{1}{x_{10}^2 s}}^z
\frac{dz'}{z'} \, \int\limits_{1/(z' s)}^{x_{10}^2} \frac{d x_{21}^2}{x_{21}^2} \: \Gamma_5 (x^2_{10},  x_{21}^2, z' s) ,
\end{align}
where 
\begin{align}
\Gamma_5 (x^2_{10},  x_{21}^2, z' s) =  \frac{\epsilon^{km} x_{10}^m}{x_{10}^2} \nabla_{10}^i \int d^2 x_1 \, x_1^k \, \Gamma_{10 \, , \, 21}^i (z' s) .
\end{align}

To simplify the remaining terms on the right of \eq{G5evol2} we replace $d^2 x_2 \, d^2 x_1 \to d^2 x_{21} \, d^2 x_2$. In Section~\ref{sec:quarkOAMevol} we have shown that 
\begin{align}\label{1st_moments}
\int d^2 x_2 \, x_2^k \, \Gamma_{20 \, , \, 21}^{gen} (z' s) \sim (z' s)^{2 \sqrt{\frac{\as N_c }{2 \pi}}}, \ \ \ \int d^2 x_2 \, x_2^k \, G_{21} (z' s) \sim (z' s)^{2 \sqrt{\frac{\as N_c }{2 \pi}}} ,
\end{align}
as follows from Eqs.~\eqref{Ikdef}, \eqref{Idef} and \eqref{Isol_full}. For small-$x$ asymptotics of quark OAM considered above this behavior was found to be subleading. Below we will proceed assuming that the expressions in \eq{1st_moments} are also subleading here, and neglect these expressions when evaluating the terms on the right of  \eq{G5evol2} containing $\Gamma_{20 \, , \, 21}^{gen}$, $G_{21}$ and $\Gamma_{21 \, , \, 20}^{gen}$. This approach will be justified by the fact that the term that would be left in the end of the calculation would scale with a higher power of energy than the terms in \eq{1st_moments}.

Neglecting the terms in \eq{1st_moments} we write
\begin{align}\label{term1}
 & \frac{\epsilon^{km} x_{10}^m}{x_{10}^2} \nabla_{10}^i 
 \int d^2 x_2 \, d^2 x_1 \, x_1^k \, \ln\frac{1}{x_{21}^2 \Lambda^2} \: \frac{ \epsilon^{i
      j} \, x_{21}^j }{ x_{21}^2 } \: \Big[ \Gamma_{20 \, , \,
    21}^{gen} (z' s) + G_{21} (z' s) \Big] \notag \\ & = \frac{\epsilon^{km} x_{10}^m}{x_{10}^2} \nabla_{10}^i 
  \int d^2 x_{21} \, d^2 x_2 \, (-x_{21}^k) \, \ln\frac{1}{x_{21}^2 \Lambda^2} \: \frac{ \epsilon^{i
      j} \, x_{21}^j }{ x_{21}^2 } \: \Big[ \Gamma_{20 \, , \,
    21}^{gen} (z' s) + G_{21} (z' s) \Big] \notag \\ & = \frac{\epsilon^{km} x_{10}^m}{x_{10}^2} \nabla_{10}^i 
  \int d^2 x_{21} \, (-x_{21}^k) \, \ln\frac{1}{x_{21}^2 \Lambda^2} \: \frac{ \epsilon^{i
      j} \, x_{21}^j }{ x_{21}^2 } \: \Big[ \Gamma^{gen} (x_{20}^2 ,  x_{21}^2 ,  z' s) + G (x^2_{21} , z' s) \Big] \notag \\ & = 
  \int d^2 x_{21} \, \frac{{\un x}_{10} \times {\un x}_{21}}{x_{10}^2}  \, \ln\frac{1}{x_{21}^2 \Lambda^2} \: \frac{ \epsilon^{i
      j} \, x_{21}^j }{ x_{21}^2 } \, \nabla_{10}^i  \: \Gamma^{gen} (x_{20}^2 ,  x_{21}^2 ,  z' s) .
\end{align}
In the last step we have noticed that the the $G$-term does not depend on $x_{10}$ and thus vanishes after differentiation. Note that
\begin{align}
\Gamma^{gen} (x_{20}^2 ,  x_{21}^2 ,  z' s) = \int d^2 x_2 \, \Gamma_{20 \, , \, 21}^{gen} (z' s)
\end{align}
where $x_{21}$ and ${\un x}_{20}$ are kept fixed during the integration.

The second (after the inhomogeneous) term on the right of \eq{G5evol2} is proportional to 
\begin{align} \label{term2}
& \frac{\epsilon^{km} x_{10}^m}{x_{10}^2} \nabla_{10}^i  \int d^2 x_2 \, d^2 x_1 \, x_1^k \, \ln\frac{1}{x_{21}^2 \Lambda^2} \: \frac{
  \epsilon^{i j} \, x_{20}^j }{ x_{20}^2 } \: \Big[
\Gamma_{20 \, , \, 21}^{gen} (z' s) + \Gamma_{21 \, , \, 20}^{gen} (z'
s) \Big] \notag \\ & = \frac{\epsilon^{km} x_{10}^m}{x_{10}^2} \nabla_{10}^i  \int d^2 x_{21} \, d^2 x_2 \, (- x_{21}^k) \, \ln\frac{1}{x_{21}^2 \Lambda^2} \: \frac{
  \epsilon^{i j} \, x_{20}^j }{ x_{20}^2 } \: \Big[
\Gamma_{20 \, , \, 21}^{gen} (z' s) + \Gamma_{21 \, , \, 20}^{gen} (z'
s) \Big] \notag \\ & = \frac{\epsilon^{km} x_{10}^m}{x_{10}^2}   \int d^2 x_{21} \, (- x_{21}^k) \, \ln\frac{1}{x_{21}^2 \Lambda^2} \: \frac{
  \epsilon^{i j} \, x_{20}^j }{ x_{20}^2 } \nabla_{10}^i \: \Big[
\Gamma^{gen} (x^2_{20}, x^2_{21} , z' s) + \Gamma^{gen} ( x^2_{21} , x^2_{20} , z'
s) \Big] = 0 
\end{align}
since 
\begin{align}
\epsilon^{i j} \, x_{20}^j \,  \nabla_{10}^i \: \Big[
\Gamma^{gen} (x^2_{20}, x^2_{21} , z' s) + \Gamma^{gen} ( x^2_{21} , x^2_{20} , z'
s) \Big] \sim \epsilon^{i j} \, x_{20}^j \, x_{20}^i =0. 
\end{align}
We again discarded the terms in \eq{1st_moments} as subleading. 

Substituting Eqs.~\eqref{term1} and \eqref{term2} into \eq{G5evol2} we arrive at
\begin{align} 
  \label{G5evol4}
  G_{5} (x_{10}^2, z s) = & \, G_{5}^{(0)} (x_{10}^2, z s) + \frac{\alpha_s
    N_c}{2\pi^2} \int\limits_{\frac{\Lambda^2}{s}}^z \frac{dz'}{z'} \, \int d^2 x_{21} \, \frac{{\un x}_{10} \times {\un x}_{21}}{x_{10}^2}  \, \ln\frac{1}{x_{21}^2 \Lambda^2} \: \frac{ \epsilon^{i
      j} \, x_{21}^j }{ x_{21}^2 } \, \nabla_{10}^i  \: \Gamma^{gen} (x_{20}^2 ,  x_{21}^2 ,  z' s) 
\notag \\ &
- \frac{\alpha_s N_c}{2\pi} \int\limits_{\frac{1}{x_{10}^2 s}}^z
\frac{dz'}{z'} \, \int\limits_{1/(z' s)}^{x_{10}^2} \frac{d x_{21}^2}{x_{21}^2} \: \Gamma_5 (x^2_{10},  x_{21}^2, z' s) ,
\end{align}
which is an equation containing two unknowns ($G_5$, $\Gamma_5$) and a known function $\Gamma^{gen} (x_{20}^2 ,  x_{21}^2 ,  z' s) $. This is also similar to the gluon helicity evolution case  \cite{Kovchegov:2017lsr}. Again the  $\Gamma^{gen} (x_{20}^2 ,  x_{21}^2 ,  z' s) $ term contains an extra $\ln s$ in the initial conditions, which makes up for the leading-logarithmic (and not DLA) structure of the kernel acting on it in \eq{G5evol4} by providing one missing logarithm of energy. 

Note that \cite{Kovchegov:2016weo,Kovchegov:2017jxc}
\begin{align}
\Gamma^{gen} (x_{20}^2 ,  x_{21}^2 ,  z' s)  \sim (z' s)^{\frac{4}{\sqrt{3}} \, \sqrt{\frac{\as
        \, N_c}{2 \pi}} } 
\end{align}
and is dominant at high energy compared to the terms in  \eq{1st_moments}, justifying us neglecting the latter.

A set of steps similar to those needed to arrive at \eq{G5evol4} when applied to \eq{M:evol4_2} gives 
\begin{align} \label{Gamma5evol3}
\Gamma_5 (x^2_{10},  x_{21}^2, z' s) = & \, G_{5}^{(0)} (x_{10}^2, z s)  +
\frac{\alpha_s N_c}{2\pi^2} \int\limits_{\frac{\Lambda^2}{s}}^{z'}
\frac{dz''}{z''} \int d^2 x_{31} \, \frac{{\un x}_{10} \times {\un x}_{31}}{x_{10}^2} \, \ln\frac{1}{x_{31}^2 \Lambda^2} \: \frac{
  \epsilon^{i j} \, x_{31}^j }{x_{31}^2} \:  \nabla_{10}^i \, \Gamma^{gen} (x^2_{30}, 
  x^2_{31} , z'' s)
\notag \\ &
- \frac{\alpha_s N_c}{2\pi} \int\limits_{\frac{1}{x_{10}^2 s}}^{z'}
\frac{dz''}{z''}   \int\limits^{\min\left[ x_{10}^2 \, , \, x_{21}^2 \tfrac{z'}{z''} \right]}_{1/(z'' s)} \frac{d x_{31}^2}{x_{31}^2} \ \Gamma_5 (x^2_{10}, x^2_{31} , z'' s)  . 
\end{align}

We thus have the following coupled system of equations:
\begin{subequations}\label{GG5}
\begin{align}
G_{5} (x_{10}^2, z s) = & \, G_{5}^{(0)} (x_{10}^2, z s) + \frac{\alpha_s
    N_c}{2\pi^2} \int\limits_{\frac{\Lambda^2}{s}}^z \frac{dz'}{z'} \, \int d^2 x_{21} \, \frac{{\un x}_{10} \times {\un x}_{21}}{x_{10}^2}  \, \ln\frac{1}{x_{21}^2 \Lambda^2} \: \frac{ \epsilon^{i
      j} \, x_{21}^j }{ x_{21}^2 } \nabla_{10}^i  \: \Gamma^{gen} (x_{20}^2 ,  x_{21}^2 ,  z' s) 
\notag \\ &
- \frac{\alpha_s N_c}{2\pi} \int\limits_{\frac{1}{x_{10}^2 s}}^z
\frac{dz'}{z'} \, \int\limits_{1/(z' s)}^{x_{10}^2} \frac{d x_{21}^2}{x_{21}^2} \: \Gamma_5 (x^2_{10},  x_{21}^2, z' s) ,  \\
 \Gamma_5 (x^2_{10},  x_{21}^2, z' s) = & \, G_{5}^{(0)} (x_{10}^2, z s)  +
\frac{\alpha_s N_c}{2\pi^2} \int\limits_{\frac{\Lambda^2}{s}}^{z'}
\frac{dz''}{z''} \int d^2 x_{31} \, \frac{{\un x}_{10} \times {\un x}_{31}}{x_{10}^2} \, \ln\frac{1}{x_{31}^2 \Lambda^2} \: \frac{
  \epsilon^{i j} \, x_{31}^j }{x_{31}^2} \:  \nabla_{10}^i \, \Gamma^{gen} (x^2_{30}, 
  x^2_{31} , z'' s)
\notag \\ &
- \frac{\alpha_s N_c}{2\pi} \int\limits_{\frac{1}{x_{10}^2 s}}^{z'}
\frac{dz''}{z''}   \int\limits^{\min\left[ x_{10}^2 \, , \, x_{21}^2 \tfrac{z'}{z''} \right]}_{1/(z'' s)} \frac{d x_{31}^2}{x_{31}^2} \ \Gamma_5 (x^2_{10}, x^2_{31} , z'' s)  .
\end{align}
\end{subequations}

Equations \eqref{GG5} have the same structure as the equations for $G_2$ and $\Gamma_2$, see Eqs. (98) in \cite{Kovchegov:2017lsr},
\begin{subequations} \label{Y:evol6}
  \begin{align} 
    \label{Y:evol6_1}
    G_2 (x_{10}^2 , z s) &= - \left(
      \frac{\alpha_s N_c}{3 \pi} \frac{1}{\alpha_h^q} G_{0} \right)
    \left( z s \, x_{10}^2 \right)^{\alpha_h^q} \, \ln\frac{1}{x_{10}^2
      \Lambda^2}
- \frac{\alpha_s N_c}{2\pi} \int\limits_{\frac{1}{x_{10}^2 s}}^z
\frac{dz'}{z'} \int\limits_{\frac{1}{z' s}}^{x_{10}^2} \frac{d
  x_{21}^2}{x_{21}^2} \, \Gamma_2 (x_{10}^2 , x_{21}^2 , z' s) ,
\\ \notag \\ \label{Y:evol6_2}
\Gamma_2 (x_{10}^2 , x_{21}^2 , z' s) &= -
\left( \frac{\alpha_s N_c}{3 \pi} \frac{1}{\alpha_h^q} G_{0} \right)
\left( z' s \, x_{10}^2 \right)^{\alpha_h^q} \, \ln\frac{1}{x_{10}^2
  \Lambda^2} 
 \\ & \hspace{3cm} - \frac{\alpha_s N_c}{2\pi} \int\limits_{\frac{1}{x_{10}^2 s}}^{z'}
\frac{dz''}{z''} \int\limits_{\frac{1}{z'' s}}^{\min\left[ x_{10}^2 \,
    , \, x_{21}^2 \tfrac{z'}{z''} \right]} \frac{dx_{31}^2}{x_{31}^2}
\, \Gamma_2 (x_{10}^2 , x_{31}^2 , z'' s) , \notag
\end{align}
\end{subequations}
where
\begin{align}
\Gamma_2 (x_{10}^2,  x_{21}^2 , z' s)  = \frac{\epsilon^{jk} x_{10}^k}{x_{10}^2}  \int d^2 x_1 \, \Gamma^j_{10, 21} (z' s). 
\end{align}

In fact, the only difference between \eqref{GG5} and \eqref{Y:evol6} is due to the inhomogeneous terms. In order to compare the two sets of equations, we have to compare their inhomogeneous terms. To do this, we can employ the exact solution of the impact parameter-integrated quark helicity equations \eqref{GGamma_eqs} found in \cite{Kovchegov:2017jxc},
\begin{subequations} \label{M:asymsol}
\begin{align}
  G(x_{10}^2 , z s) &= \frac{1}{3} G_0 \: (z s \, x_{10}^2)^{\alpha_h^q} \\
  \Gamma(x_{10}^2 , x_{21}^2 , z s) &= \frac{1}{3} G_0 \: (z s \,
  x_{21}^2)^{\alpha_h^q} \left[ 4
    \left(\frac{x_{10}^2}{x_{21}^2}\right)^{\frac{\alpha_h^q}{4}} - 3
  \right] ,
\end{align}
\end{subequations}
where, as in the above, the ``quark helicity intercept'' is given by
\begin{align} 
  \label{M:ahel}
  \alpha_h^q = \frac{4}{\sqrt{3}} \sqrt{\frac{\alpha_s N_c}{2\pi}}
  \approx 2.31 \sqrt{\frac{\alpha_s N_c}{2\pi}} .
\end{align}
$G_0$ is the inhomogeneous term in the impact parameter-integrated version of Eqs.~\eqref{GGamma_eqs}, assumed for simplicity to be constant in \cite{Kovchegov:2017jxc}, and
\begin{align}
\Gamma (x_{10}^2 ,  x_{21}^2 ,  z' s) = \int d^2 x_2 \, \Gamma_{10 \, , \, 21} (z' s).
\end{align}

After a straightforward differentiation we arrive at the following expression for the second inhomogeneous term in  \eqref{GG5}: 
\begin{align}
& \frac{\alpha_s
    N_c}{2\pi^2} \int\limits_{\frac{\Lambda^2}{s}}^z \frac{dz'}{z'} \, \int d^2 x_{21} \, \frac{{\un x}_{10} \times {\un x}_{21}}{x_{10}^2}  \, \ln\frac{1}{x_{21}^2 \Lambda^2} \: \frac{ \epsilon^{i
      j} \, (x_{21})_\bot^j }{ x_{21}^2 } \nabla_{10}^i  \: \Gamma^{gen} (x_{20}^2 ,  x_{21}^2 ,  z' s)  \\ & =  \frac{\alpha_s
    N_c}{\pi^2} \, \alpha_h^q \, \frac{G_0}{3} \, \int\limits_{\frac{\Lambda^2}{s}}^z \frac{dz'}{z'} \, (z' s x_{10}^2)^{\alpha_h^q} \, K  \left( \frac{1}{x_{10}^2 \Lambda^2} \right) , 
\end{align}
where we have defined
\begin{align}\label{K-def}   
 K  \left( \frac{1}{x_{10}^2 \Lambda^2} \right)  \equiv   \int d^2 x_{21} \, \frac{({\un x}_{10} \times {\un x}_{21})^2}{x_{10}^2 \, x_{21}^2 \, x_{20}^2}  \, \ln\frac{1}{x_{21}^2 \Lambda^2} \left[ \theta (x_{20} - x_{21}) \, \left(\frac{x_{21}^2}{x_{10}^2}\right)^{\alpha_h^q} \, \left(\frac{x_{20}^2}{x_{21}^2}\right)^{\frac{\alpha_h^q}{4}} + \theta (x_{21} - x_{20}) \, \left(\frac{x_{20}^2}{x_{10}^2}\right)^{\alpha_h^q}  \right]. 
\end{align}
The integral in \eq{K-def} is IR-divergent. If we cut it off by $1/\Lambda$ in the IR, and put $\alpha_h^q =0$ in it (thus neglecting higher powers of $\as$ not enhanced by large logarithms of energy), we get
\begin{align}
K  \left( \frac{1}{x_{10}^2 \Lambda^2} \right)  \approx \frac{\pi}{4} \, \ln^2 \left( \frac{1}{x_{10}^2 \Lambda^2} \right). 
\end{align}
A more careful evaluation of \eq{K-def}, neglecting $\alpha_h^q$ only compared to order-one constants, yields
\begin{align}\label{Kfull}
K \left( \frac{1}{x_{10}^2 \Lambda^2} \right) \approx \frac{\pi}{2} \, \left( 1 - \ln \frac{1}{x_{10}^2 \Lambda^2} \right) +  \frac{\pi}{2 \, (\alpha_h^q)^2} \, \left[ \left( \frac{1}{x_{10}^2 \Lambda^2} \right)^{\alpha_h^q}  - 1 - \alpha_h^q \, \ln \frac{1}{x_{10}^2 \Lambda^2} \right].
\end{align}
The evolution equations for $G_5$ and $\Gamma_5$ become
\begin{subequations}\label{GG7}
\begin{align}
& G_{5} (x_{10}^2, z s) = G_{5}^{(0)} (x_{10}^2, z s) + \frac{\alpha_s N_c}{\pi^2} \, \frac{G_0}{3} \, K \left( \frac{1}{x_{10}^2 \Lambda^2} \right)  \, (z s x_{10}^2)^{\alpha_h^q}  
- \frac{\alpha_s N_c}{2\pi} \int\limits_{\frac{1}{x_{10}^2 s}}^z
\frac{dz'}{z'} \, \int\limits_{1/(z' s)}^{x_{10}^2} \frac{d x_{21}^2}{x_{21}^2} \: \Gamma_5 (x^2_{10},  x_{21}^2, z' s) ,  \\
 & \Gamma_5 (x^2_{10},  x_{21}^2, z' s)  = G_{5}^{(0)} (x_{10}^2, z' s) + \frac{\alpha_s N_c}{\pi^2} \, \frac{G_0}{3} \, K \left( \frac{1}{x_{10}^2 \Lambda^2} \right) \, (z' s x_{10}^2)^{\alpha_h^q} 
\\ & \hspace*{5cm} - \frac{\alpha_s N_c}{2\pi} \int\limits_{\frac{1}{x_{10}^2 s}}^{z'}
\frac{dz''}{z''}   \int\limits^{\min\left[ x_{10}^2 \, , \, x_{21}^2 \tfrac{z'}{z''} \right]}_{1/(z'' s)} \frac{d x_{31}^2}{x_{31}^2} \ \Gamma_5 (x^2_{10}, x^2_{31} , z'' s)  . \notag
\end{align}
\end{subequations}
Note that, as follows from \eq{G10init} above, 
\begin{align}\label{G50}
G_{5}^{(0)} (x_{10}^2, z s) = G_{2}^{(0)} (x_{10}^2, z s) = - \frac{\as^2 \, C_F}{N_c} \, \pi \, \ln \left( \frac{1}{x_{10} \, \Lambda} \right). 
\end{align}

Equations \eqref{GG7} have two inhomogeneous terms. The following calculation would be simplified if we could neglect one compared to the other. To see which one to neglect, let us first do some power counting. The initial condition for the quark helicity evolution is of the order $G_0 \sim \as^2 \, \ln s \sim \as^{3/2}$  \cite{Kovchegov:2017lsr} if we assume that $\ln s \sim 1/\sqrt{\as}$, as is appropriate for the DLA limit. We thus have, for the two inhomogeneous terms in Eqs.~\eqref{GG7},
\begin{align}
G_{5}^{(0)} \sim \as^2 \gg  \as \, G_0 \, K \sim \as^{5/2}.
\end{align}
It appears that we can neglect the second inhomogeneous term compared to the first one. (Note that the situation here is slightly different from the equations for gluon helicity obtained in \cite{Kovchegov:2017lsr}, where the second inhomogeneous term is, parametrically, of the same order as the first term, $\sqrt{\as} \, G_0 \sim \as^2 \sim G_{2}^{(0)}$, and the first term is neglected due to the lack of power of $ z s$ enhancement at Born level.) However, since equations \eqref{GG7} are linear, their solution is a sum of solutions of the same equations with one set of equations having only the $G_{5}^{(0)}$ inhomogeneous terms, while another one containing only the $ \as \, G_0 \, K$ inhomogeneous terms. 

To find the solution of the former equations, keep only the $G_5^{(0)}$ inhomogeneous terms in Eqs.~\eqref{GG7} and substitute \eq{G50} into Eqs.~\eqref{GG7}. We arrive at
\begin{subequations}\label{GG8}
\begin{align}
G_{5} (x_{10}^2, z s) &= - \frac{\as^2 \, C_F}{N_c} \, \pi \, \ln \left( \frac{1}{x_{10} \, \Lambda} \right) - \frac{\alpha_s N_c}{2\pi} \int\limits_{\frac{1}{x_{10}^2 s}}^z
\frac{dz'}{z'} \, \int\limits_{1/(z' s)}^{x_{10}^2} \frac{d x_{21}^2}{x_{21}^2} \: \Gamma_5 (x^2_{10},  x_{21}^2, z' s) ,  \\
 \Gamma_5 (x^2_{10},  x_{21}^2, z' s) & = - \frac{\as^2 \, C_F}{N_c} \, \pi \, \ln \left( \frac{1}{x_{10} \, \Lambda} \right) - \frac{\alpha_s N_c}{2\pi} \int\limits_{\frac{1}{x_{10}^2 s}}^{z'}
\frac{dz''}{z''}   \int\limits^{\min\left[ x_{10}^2 \, , \, x_{21}^2 \tfrac{z'}{z''} \right]}_{1/(z'' s)} \frac{d x_{31}^2}{x_{31}^2} \ \Gamma_5 (x^2_{10}, x^2_{31} , z'' s)  . 
\end{align}
\end{subequations}

Defining 
\begin{align}
{\bar G}_5 (x_{10}^2, z s) = - \frac{G_{5} (x_{10}^2, z s) }{\frac{\as^2 \, C_F}{N_c} \, \pi \, \ln \left( \frac{1}{x_{10} \, \Lambda} \right)}, \ \ \ {\bar \Gamma}_5 (x^2_{10},  x_{21}^2, z' s) = - \frac{\Gamma_5 (x^2_{10},  x_{21}^2, z' s)}{\frac{\as^2 \, C_F}{N_c} \, \pi \, \ln \left( \frac{1}{x_{10} \, \Lambda} \right)}
\end{align}
we reduce Eqs.~\eqref{GG8} to 
\begin{subequations}\label{GG9}
\begin{align}
{\bar G}_{5} (x_{10}^2, z s) &= 1 - \frac{\alpha_s N_c}{2\pi} \int\limits_{\frac{1}{x_{10}^2 s}}^z
\frac{dz'}{z'} \, \int\limits_{1/(z' s)}^{x_{10}^2} \frac{d x_{21}^2}{x_{21}^2} \: {\bar \Gamma}_5 (x^2_{10},  x_{21}^2, z' s) ,  \\
{\bar \Gamma}_5 (x^2_{10},  x_{21}^2, z' s) & = 1 - \frac{\alpha_s N_c}{2\pi} \int\limits_{\frac{1}{x_{10}^2 s}}^{z'}
\frac{dz''}{z''}   \int\limits^{\min\left[ x_{10}^2 \, , \, x_{21}^2 \tfrac{z'}{z''} \right]}_{1/(z'' s)} \frac{d x_{31}^2}{x_{31}^2} \ {\bar \Gamma}_5 (x^2_{10}, x^2_{31} , z'' s)  ,
\end{align}
\end{subequations}
which are solved in Appendix \ref{app:Born} (for $\beta = -1$ there) with the solution for ${\bar G}_{5}$ given by \eq{oscill0}. Employing the latter we write
\begin{align}\label{oscill}
G_5 (s_{10}, \eta) = - \frac{\as^2 \, C_F}{N_c} \, \pi \, \ln \left( \frac{1}{x_{10} \, \Lambda} \right) \, \frac{J_1 \left( \sqrt{\frac{2 \alpha_s N_c}{\pi}} \,  \ln (z  s \, x_{10}^2) \right)}{\sqrt{\frac{\alpha_s N_c}{2\pi}} \, \ln (z  s \, x_{10}^2)} .
\end{align}
This result also oscillates with a decreasing amplitude for increasing $z  s \, x_{10}^2$. As we will shortly see, it is negligible compared to the solution of Eqs.~\eqref{GG7} with the $ \as \, G_0 \, K$ inhomogeneous terms. 

Keeping only the $ \as \, G_0 \, K$ inhomogeneous terms in Eqs.~\eqref{GG7} we have
\begin{subequations}\label{GG6}
\begin{align}
G_{5} (x_{10}^2, z s) &= \frac{\alpha_s N_c}{\pi^2} \, \frac{G_0}{3} \, K \left( \frac{1}{x_{10}^2 \Lambda^2} \right)  \, (z s x_{10}^2)^{\alpha_h^q}  
- \frac{\alpha_s N_c}{2\pi} \int\limits_{\frac{1}{x_{10}^2 s}}^z
\frac{dz'}{z'} \, \int\limits_{1/(z' s)}^{x_{10}^2} \frac{d x_{21}^2}{x_{21}^2} \: \Gamma_5 (x^2_{10},  x_{21}^2, z' s) ,  \\
 \Gamma_5 (x^2_{10},  x_{21}^2, z' s) & =  \frac{\alpha_s N_c}{\pi^2} \, \frac{G_0}{3} \, K \left( \frac{1}{x_{10}^2 \Lambda^2} \right) \, (z' s x_{10}^2)^{\alpha_h^q} 
- \frac{\alpha_s N_c}{2\pi} \int\limits_{\frac{1}{x_{10}^2 s}}^{z'}
\frac{dz''}{z''}   \int\limits^{\min\left[ x_{10}^2 \, , \, x_{21}^2 \tfrac{z'}{z''} \right]}_{1/(z'' s)} \frac{d x_{31}^2}{x_{31}^2} \ \Gamma_5 (x^2_{10}, x^2_{31} , z'' s)  . 
\end{align}
\end{subequations}
Comparing \eqref{GG6} to \eqref{Y:evol6} we conclude that 
\begin{align}\label{G52gen}
G_{5} (x_{10}^2, z s) = - \frac{\alpha_h^q \, K \left( \frac{1}{x_{10}^2 \Lambda^2} \right)}{\pi  \, \ln \left( \frac{1}{x_{10}^2 \Lambda^2} \right)} \, G_2 (x_{10}^2 , z s) . 
\end{align}
Since, as was shown in \cite{Kovchegov:2017lsr},
\begin{align}
G_2 (x_{10}^2 , z s) \sim (z s)^{\frac{13}{4 \sqrt{3}}\sqrt{\frac{\alpha_s N_c} {2\pi}}},
\end{align} 
we conclude that 
\begin{align}
G_5 (x_{10}^2 , z s) \sim (z s)^{\frac{13}{4 \sqrt{3}}\sqrt{\frac{\alpha_s N_c} {2\pi}}},
\end{align} 
and the solution of Eqs.~\eqref{GG6} dominates over \eq{oscill} at high energy. Hence, while the $G_{5}^{(0)}$ inhomogeneous terms in Eqs.~\eqref{GG7} are parametrically larger than the $ \as \, G_0 \, K$ inhomogeneous terms, we are justified to only keep the latter, since at high energies they give the dominant contribution. Therefore, \eq{G52gen} gives us the leading high-energy solution of Eqs.~\eqref{GG7} for $G_5$.

%
\subsection{Gluon OAM distribution at small $x$} \label{sec:result}
%

Using \eq{G52gen} in \eq{OAM20} we obtain
\begin{align}\label{OAM21}
L_G (x, Q^2)  =  \frac{8 i N_c}{g^2 \, (2\pi)^3} \,  \int d^2 x_{10} \,  d^2 k_\perp \, 
  e^{i \un{k} \cdot {\un x}_{10}} \, \left( \un{k} \cdot {\un x}_{10} \right)  
 \,  \frac{\alpha_h^q \, K \left( \frac{1}{x_{10}^2 \Lambda^2} \right)}{\pi  \, \ln \left( \frac{1}{x_{10}^2 \Lambda^2} \right)} \, G_{2} \left( x_{10}^2, s \right).
\end{align} 
This has to be compared with the gluon helicity distribution (see \eqref{Ghel_TMD}) 
\begin{align}\label{Ghel2}
\Delta G (x, Q^2) = \frac{8 i N_c}{g^2 \, (2\pi)^3} \,  \int d^2 x_{10} \, d^2 k_\perp \, 
  e^{i \un{k} \cdot {\un x}_{10}} \, \left( \un{k} \cdot {\un x}_{10} \right)  
 \, G_{2} \left( x_{10}^2, s \right) .
\end{align}

Assuming that, after all integrations are carried out, at the leading DLA level we can simply replace
\begin{align}
\frac{1}{x_{10}^2}   \to Q^2
\end{align}
with $Q^2$ being the upper cutoff on the $k_T$ integral in both \eqref{OAM21} and \eqref{Ghel2}, we arrive at 
\begin{align}\label{LGDG3}
L_G (x, Q^2)  = \frac{\alpha_h^q \,  K \left( \frac{Q^2}{\Lambda^2} \right)}{\pi \, \ln \left( \frac{Q^2}{\Lambda^2} \right)}   \, \Delta G (x, Q^2) .
\end{align}

Noticing that the first term on the right-hand side of \eq{Kfull} can be thought of as a constant under one of the logarithms  in the $\alpha_h^q$-expansion of the second term, we keep only this second term on the right-hand side of \eqref{Kfull} to write
\begin{align}\label{LGDG4}
L_G (x, Q^2)  = \frac{\left( \frac{Q^2}{\Lambda^2} \right)^{\alpha_h^q}  - 1 - \alpha_h^q \, \ln \frac{Q^2}{\Lambda^2}}{2 \, \alpha_h^q \, \ln \left( \frac{Q^2}{\Lambda^2} \right)}   \, \Delta G (x, Q^2) .
\end{align}
Unlike the quark OAM in \eq{e:MAINRESULT2}, this result appears to be different from the DGLAP-based conclusion reached in \cite{Hatta:2018itc}. However, the conclusion in \cite{Hatta:2018itc} was based on the assumption that $|\Delta G| \gg |\Delta q|$, which is the opposite of what was found in the DLA approximation \cite{Kovchegov:2016weo,Kovchegov:2017jxc,Kovchegov:2017lsr}. 

The prefactor in the relation \eqref{LGDG4} resums powers of $\as \, \ln^2 \frac{Q^2}{\Lambda^2}$, which are strictly-speaking are not DLA. (In the DLA approximation one only keeps powers of $\as \, \ln^2 (1/x)$.) Therefore, it is possible that one has to expand \eq{LGDG4} to the lowest non-trivial order in $\alpha_h^q$, obtaining 
\begin{align}\label{LGDG2}
L_G (x, Q^2)  = \left( \frac{\alpha_h^q}{4} \, \ln \frac{Q^2}{\Lambda^2} \right) \, \Delta G (x, Q^2) . 
\end{align}

The small-$x$ asymptotics of the gluon OAM easily follows from either \eq{LGDG4} or \eq{LGDG2} using the DLA asymptotics of the gluon helicity distribution found in \cite{Kovchegov:2017lsr}. We conclude that
\begin{align} \label{e:MAINRESULT}
L_G (x, Q^2)  \sim \Delta G (x, Q^2) \sim \left(\frac{1}{x}\right)^{\alpha_h^G} \sim
  \left(\frac{1}{x}\right)^{\frac{13}{4 \sqrt{3}} \, \sqrt{\frac{\as
        \, N_c}{2 \pi}}} \sim \left(\frac{1}{x}\right)^{1.88 \,
    \sqrt{\frac{\as \, N_c}{2 \pi}}}.
\end{align}
In Appendix~\ref{app:large} we present a simple toy model describing a way of thinking about the DLA evolution for gluon OAM changing its relation to the gluon helicity from $L_G = - \Delta G$ to \eq{LGDG2}.

%
\section{Summary} \label{sec:sum}
%

The calculation performed in this paper heavily relied on the earlier works \cite{Kovchegov:2015pbl,Kovchegov:2016weo,Kovchegov:2016zex,Kovchegov:2017jxc,Kovchegov:2017lsr,Kovchegov:2018znm}. By simplifying the quark and gluon OAM distributions definitions at small $x$ to Eqs.~\eqref{qOAM14.5} and  \eqref{OAM19} we managed to relate these quantities to the polarized dipole amplitudes $G_{10} (z s)$ and $G_{10}^i (z s)$ respectively, which were employed previously for determination of the quark and gluon helicity distributions. The small-$x$ asymptotics of $L_{q + \bar{q}}  (x, Q^2)$ and $L_G (x, Q^2)$ should have resulted from the high-energy asymptotics of these quantities. Naively, the latter could be easily found from the solutions of the DLA evolution equations for $G_{10} (z s)$ and $G_{10}^i (z s)$ constructed in \cite{Kovchegov:2016weo,Kovchegov:2017jxc,Kovchegov:2017lsr}. However, one had to be careful here, since in \cite{Kovchegov:2016weo,Kovchegov:2017jxc,Kovchegov:2017lsr} we found the expressions for $G_{10} (z s)$ and $G_{10}^i (z s)$ integrated over all impact parameters. At the same time, Eqs.~\eqref{qOAM14.5} and  \eqref{OAM19} contain $G_{10} (z s)$ and $G_{10}^i (z s)$ respectively, weighted by the position of the polarized quark ${\un x}_1$, and then integrated over all impact parameters. So an analysis of the first ${\un x}_1$-moments of $G_{10} (z s)$ and $G_{10}^i (z s)$ was in order. 

In the quark OAM case, the first ${\un x}_1$-moment of $G_{10} (z s)$ turned out to be subleading at small $x$, such that \eq{qOAM14.5}, after we discarded the first term on its right (which was proportional to the first moment), led to (cf. \cite{Hatta:2018itc}) 
\begin{align} \label{e:MAINRESULT2copy}
 L_{q + \bar{q}} (x, Q^2)  = - \Delta \Sigma (x, Q^2) \sim 
  \left(\frac{1}{x}\right)^{\frac{4}{\sqrt{3}} \, \sqrt{\frac{\as
        \, N_c}{2 \pi}} } .
\end{align}

The gluon OAM distribution in \eq{OAM19} is directly proportional to the first ${\un x}_1$-moment of $G_{10}^i (z s)$. Constructing the small-$x$ asymptotics of this moment we arrived at 
\begin{align}\label{LGDG30}
L_G (x, Q^2)  = \left( \frac{\alpha_h^q}{4} \, \ln \frac{Q^2}{\Lambda^2} \right) \, \Delta G (x, Q^2) \sim
  \left(\frac{1}{x}\right)^{\frac{13}{4 \sqrt{3}} \, \sqrt{\frac{\as
        \, N_c}{2 \pi}}}. 
\end{align}
Equations \eqref{e:MAINRESULT2copy} and \eqref{LGDG30} summarize our main results in this paper.

One may be concerned about an apparent asymmetry: the first ${\un x}_1$-moment of $G_{10} (z s)$ is subleading at small $x$, while the first ${\un x}_1$-moment of $G_{10}^i (z s)$, labeled $G_5$ above, appears not to be subleading, and leads to \eq{LGDG30}. Note, however, that in the strict DLA analysis the prefactor on the right of \eq{LGDG30} is subleading, $\sim \alpha_h^q \, \ln (Q^2/\Lambda^2) \sim \sqrt{\as} \, \ln (Q^2/\Lambda^2) \ll 1$. Therefore, the first moment $G_5$, and, consequently, $L_G (x, Q^2)$ at small $x$ are also subleading, by apparent analogy to the quark OAM case. This is probably the consequence of keeping only the parametrically subleading but dominant at high energy inhomogeneous term $\as \, G_0 \, K$ in Eqs.~\eqref{GG7} to arrive at the solution \eqref{G52gen} for $G_5$.  The difference in the gluon OAM case as compared to the quark OAM is that, while subleading in DLA, the $G_5$ term gives us the only contribution to $L_G (x, Q^2)$ and cannot be neglected. One may even speculate that our conclusion \eqref{LGDG30} could be more conservatively formulated as $|L_G| \ll |\Delta G|$ at small $x$.


\section*{Acknowledgments}

Special thanks go to Feng Yuan for suggesting that the author look at quark and gluon OAM at small $x$ and to Yoshitaka Hatta for many critical discussions of this work. The author would also like to thank Matt Sievert for his interest in this project. 

This material is based upon work supported by the U.S. Department of
Energy, Office of Science, Office of Nuclear Physics under Award
Number DE-SC0004286. The author thanks the U.S. Department of Energy's Institute for Nuclear Theory at the University of Washington for its hospitality and support during the final stages of this work (INT pre-print number INT-PUB-18-065).  \\



\appendix
\section{The saturation/CGC averaging}
\label{app:CGCave}

Start with the an expectation value of some operator $\hat{\cal O} (b,r)$ in the proton state $\ket{P}$. The expectation value $\bra{P} \hat{\cal O} (b,r) \ket{P}$ is independent of $b = (b^-, {\un b})$. It should be proportional to the CGC-averaged operator, integrated over all space:
\begin{align}
\bra{P} \hat{\cal O} (b,r) \ket{P} \sim \int d^2 b \, d b^- \, \left\langle \hat{\cal O} (b,r) \right\rangle .
\end{align}
To fix the normalization we put $\hat{\cal O} (b,r) = 1$ and note that $\braket{P}{P} = 2 P^+ \, (2 \pi)^3 \, \delta (0^-) \, \delta^2 (\un{0}) = 2 P^+ V^-$ with $V^- = \int d^2 x \, d x^-$. We get
\begin{align}\label{matrix_el1}
\bra{P} \hat{\cal O} (b,r) \ket{P} = 2 P^+ \,  \int d^2 b \, d b^- \, \left\langle \hat{\cal O} (b,r) \right\rangle .
\end{align}

Now, since $\bra{P} \hat{\cal O} (b,r) \ket{P} = \bra{P} \hat{\cal O} (0,r) \ket{P}$, the off-forward matrix element is a Fourier transform of the CGC-averaged operator, with the normalization fixed by \eq{matrix_el1},
\begin{align}\label{matrix_el2}
\bra{P+ \frac{\Delta}{2}} \hat{\cal O} (0,r) \ket{P - \frac{\Delta}{2}} = 2 P^+ \,  \int d^2 b \, d b^- \, e^{- i b \cdot \Delta} \, \left\langle \hat{\cal O} (b,r) \right\rangle ,
\end{align}
where $\Delta = (\Delta^+, \un{\Delta})$. (The sign in the Fourier exponent is due to $i (0-b) \cdot \Delta = - i b \cdot \Delta$.) Inverting this Fourier transform we arrive at
\begin{align}\label{matrix_el3}
\left\langle \hat{\cal O} (b,r) \right\rangle = \frac{1}{2 P^+} \, \int \frac{d^2 \Delta \, d \Delta^+}{(2 \pi)^3} \, e^{i b \cdot \Delta} \, \bra{P+ \frac{\Delta}{2}} \hat{\cal O} (0,r) \ket{P - \frac{\Delta}{2}} .
\end{align}


\section{Solution of a useful system of integral equations}

\label{app:Born}

In the main text, on two separate occasions, we needed to solve the following system of integral questions
\begin{subequations}\label{GG99}
\begin{align}
{\bar G}_{5} (x_{10}^2, z s) &= 1 + \beta \,  \frac{\alpha_s N_c}{2\pi} \int\limits_{\frac{1}{x_{10}^2 s}}^z
\frac{dz'}{z'} \, \int\limits_{1/(z' s)}^{x_{10}^2} \frac{d x_{21}^2}{x_{21}^2} \: {\bar \Gamma}_5 (x^2_{10},  x_{21}^2, z' s) ,  \\
{\bar \Gamma}_5 (x^2_{10},  x_{21}^2, z' s) & = 1 + \beta \, \frac{\alpha_s N_c}{2\pi} \int\limits_{\frac{1}{x_{10}^2 s}}^{z'}
\frac{dz''}{z''}   \int\limits^{\min\left[ x_{10}^2 \, , \, x_{21}^2 \tfrac{z'}{z''} \right]}_{1/(z'' s)} \frac{d x_{31}^2}{x_{31}^2} \ {\bar \Gamma}_5 (x^2_{10}, x^2_{31} , z'' s)  ,
\end{align}
\end{subequations}
with Eqs.~\eqref{Ievol4} corresponding to the $\beta=+1$ case and Eqs.~\eqref{GG9} corresponding to the $\beta = - 1$ case. 

Introducing the scaled logarithmic variables \cite{Kovchegov:2016weo}
\begin{subequations} \label{e:logunits}
\begin{align}
  \eta & \equiv \sqrt{\frac{\alpha_s N_c}{2\pi}} \ln\frac{z
    s}{\Lambda^2} , & \hspace{0.5cm} s_{10} & \equiv
  \sqrt{\frac{\alpha_s N_c}{2\pi}} \ln\frac{1}{x_{10}^2 \Lambda^2} ,
\\
\eta' & \equiv \sqrt{\frac{\alpha_s N_c}{2\pi}} \ln\frac{z' s}{\Lambda^2} ,
& \hspace{0.5cm}
s_{21} & \equiv \sqrt{\frac{\alpha_s N_c}{2\pi}} \ln\frac{1}{x_{21}^2 \Lambda^2} ,
\\
\eta'' & \equiv \sqrt{\frac{\alpha_s N_c}{2\pi}} \ln\frac{z'' s}{\Lambda^2} ,
& \hspace{0.5cm}
s_{32} & \equiv \sqrt{\frac{\alpha_s N_c}{2\pi}} \ln\frac{1}{x_{32}^2 \Lambda^2} ,
\end{align}
\end{subequations}
we get
\begin{subequations}\label{GGeqns}
\begin{align} \label{e:G_redef} 
  & {\bar G}_5 (s_{10}, \eta) = 1 + \beta
  \int\limits_{s_{10}}^\eta d \eta' \int\limits_{s_{10}}^{\eta'} d
  s_{21} \, {\bar \Gamma}_5 (s_{10}, s_{21}, \eta') ,
\\ \label{e:Gam_redef}
& {\bar \Gamma}_5 (s_{10}, s_{21}, \eta') = 1 + \beta 
\int\limits_{s_{10}}^{\eta'} d \eta'' \hspace{-0.4cm}
\int\limits_{\mbox{max} \left\{ s_{10}, s_{21} + \eta'' - \eta'
  \right\}}^{\eta''} \hspace*{-1cm} d s_{32}
\, {\bar \Gamma}_5 (s_{10}, s_{32}, \eta'') . 
\end{align}
\end{subequations}

By analogy to \cite{Kovchegov:2017jxc} we assume a scaling ansatz of the solution,
\begin{align}
{\bar G}_5 (s_{10}, \eta) & \, =   {\bar G}_5 ( \eta - s_{10}) , \\
{\bar \Gamma}_5 (s_{10}, s_{21}, \eta') & \, = {\bar \Gamma}_5 (\eta' - s_{10}, \eta' - s_{21}) .
\end{align}
Equations \eqref{GGeqns} become
\begin{subequations}\label{GGeqns2}
\begin{align} \label{e:G_redef2}
  {\bar G}_5 (\zeta) & = 1 + \beta \int\limits_{0}^\zeta d \xi \int\limits_{0}^{\xi}
  d \xi' \, {\bar \Gamma}_5 (\xi, \xi')  ,
\\ 
{\bar \Gamma}_5 (\zeta, \zeta') & = 1 + \beta  \int\limits_{0}^{\zeta'} d \xi
\int\limits_{0}^{\xi} d \xi' \, {\bar \Gamma}_5 (\xi, \xi') + \beta \int\limits_{\zeta'}^{\zeta} d \xi  \, \int\limits_{0}^{\zeta'} d \xi' 
{\bar \Gamma}_5 (\xi, \xi')  ,
\end{align}
\end{subequations}

Following \cite{Kovchegov:2017jxc} we write Eqs.~\eqref{GGeqns2} in a differential form,
\begin{subequations}\label{GGeqns3}
\begin{align} \label{e:G_redef3}
&  \pd_\zeta {\bar G}_5 (\zeta) =  \beta \int\limits_{0}^\zeta d \xi' \, {\bar \Gamma}_5 (\zeta, \xi')  ,
\\ 
& \pd_\zeta {\bar \Gamma}_5 (\zeta, \zeta') =  \beta  \int\limits_{0}^{\zeta'} d \xi' \, {\bar \Gamma}_5 (\zeta, \xi') ,
\end{align}
\end{subequations}
with the initial conditions
\begin{align}
{\bar G}_5 (0) = 1, \ \ \ {\bar \Gamma}_5 (\xi, \xi) = {\bar G}_5 (\xi) . 
\end{align}
These equations can be solved with the help of a Laplace transform, leading to
\begin{subequations}\label{GG5sol1}
\begin{align}
& {\bar \Gamma}_5 (\zeta, \zeta') = \int \frac{d \omega}{2 \pi i} \, e^{\omega \, \zeta' + \frac{\beta \, \zeta}{\omega}} \, \Gamma_{5\omega} (0), \\ 
& {\bar G}_5 (\zeta) = {\bar \Gamma}_5 (\zeta, \zeta) =  \int \frac{d \omega}{2 \pi i} \, e^{\left( \omega + \frac{\beta}{\omega} \right)  \, \zeta} \, \Gamma_{5\omega} (0)
\end{align}
\end{subequations}
with the still unknown function $\Gamma_{5\omega} (0)$ satisfying the following relations:
\begin{align}\label{constraints}
& \int \frac{d \omega}{2 \pi i} \, e^{\frac{\beta \, \zeta}{\omega}} \, \frac{1}{\omega} \, \Gamma_{5\omega} (0) = 0 , 
& \int \frac{d \omega}{2 \pi i} \, e^{\left( \omega + \frac{\beta}{\omega} \right)  \, \zeta} \,\omega \, \Gamma_{5\omega} (0) =0. 
\end{align}

Searching for $\Gamma_{5\omega} (0)$ in the form
\begin{align}
\Gamma_{5\omega} (0) = \sum_{n=-\infty}^{n=\infty} \, c_n \, \omega^n
\end{align}
and satisfying the constraints \eqref{constraints} we arrive at
\begin{align}\label{Gam5sol}
\Gamma_{5\omega} (0) = c_{-1} \, \left[ \frac{1}{\omega} - \frac{\beta}{\omega^3} \right] .
\end{align}
Inserting \eq{Gam5sol} into Eqs.~\eqref{GG5sol1} yields
\begin{subequations}\label{GG5sol2}
\begin{align}
& {\bar G}_5 (\zeta) =  \int \frac{d \omega}{2 \pi i} \, e^{\left( \omega + \frac{\beta}{\omega} \right)  \, \zeta} \,  \left[ \frac{1}{\omega} - \frac{\beta}{\omega^3} \right] , \label{G5sol2} \\
& {\bar \Gamma}_5 (\zeta, \zeta') = \int \frac{d \omega}{2 \pi i} \, e^{\omega \, \zeta' + \frac{\beta \, \zeta}{\omega}} \,  \left[ \frac{1}{\omega} - \frac{\beta}{\omega^3} \right] , 
\end{align}
\end{subequations}
where we have fixed $c_{-1} = 1$ by imposing the ${\bar G}_5 (0) = 1$ condition. 

Since we are interested in ${\bar G}_5 (\zeta)$, we perform the $\omega$-integral in \eq{G5sol2}, arriving at
\begin{align}\label{G5sol3}
{\bar G}_5 (\zeta) = \frac{I_1 (2 \, \zeta \, \sqrt{\beta})}{\zeta \, \sqrt{\beta}} .
\end{align}

For $\beta = +1$, \eq{G5sol3} yields
\begin{align}\label{Isol10}
{\bar I} (s_{10}, \eta) = {\bar G}_5 (s_{10}, \eta) = \frac{I_1 (2 \, (\eta - s_{10}))}{\eta - s_{10}} = \frac{I_1 \left( 2 \, \sqrt{ \frac{\alpha_s N_c}{2 \pi}} \,  \ln (z  s \, x_{10}^2) \right)}{\sqrt{\frac{\alpha_s N_c}{2\pi}} \, \ln (z  s \, x_{10}^2)} . 
\end{align}

For $\beta = -1$, \eq{G5sol3} gives
\begin{align}\label{oscill0}
{\bar G}_5 (s_{10}, \eta) = \frac{J_1 (2 \, (\eta - s_{10}))}{\eta - s_{10}} = \frac{J_1 \left( \sqrt{\frac{2 \alpha_s N_c}{\pi}} \,  \ln (z  s \, x_{10}^2) \right)}{\sqrt{\frac{\alpha_s N_c}{2\pi}} \, \ln (z  s \, x_{10}^2)} . 
\end{align}

%
\section{Comparison with the earlier works} \label{sec:comparison}
%

Here we demonstrate that our definition \eqref{OAM2} of the gluon OAM agrees with that in \cite{Jaffe:1989jz,Hatta:2016aoc}. Therefore, we are using the Jaffe-Manohar definition of the gluon OAM. Our strategy is to show that the gluon OAM definitions in \cite{Jaffe:1989jz,Hatta:2016aoc} are equivalent to each other, and that our definition is equivalent to \cite{Hatta:2016aoc}, and, hence, to \cite{Jaffe:1989jz}. 

Begin with Eq.~(4) in \cite{Hatta:2016aoc}, which we can write as follows:
\begin{align}\label{LGcomp1}
L_G & = - \frac{i \, \epsilon^{ij}}{2 S^+} \, \lim_{\Delta \to 0} \frac{\delta}{\delta \Delta^j} \, \bra{P', S} F^{+\alpha} (0) \, \overleftrightarrow{D}_\mathrm{pure}^i \, A^\mathrm{phys}_\alpha (0) \, \ket{P, S} \\ \notag & = \frac{\epsilon^{ij}}{2 S^+ \, V^-} \int d^2 x \, d x^- x^j \bra{P, S} F^{+\alpha} (x) \, \overleftrightarrow{D}_\mathrm{pure}^i \, A^\mathrm{phys}_\alpha (x) \, \ket{P, S} ,
\end{align}
where $V^- = \int d^2 x \, d x^- = (2 \pi)^3 \, \delta^2 ({\un 0}) \, \delta (0^+)$ and $\Delta^\mu = P'^\mu - P^\mu$. 

For simplicity, let us work in the $A^+ =0$ gauge with the ${\un \nabla} \cdot {\un A} (x^- = - \infty) =0$ sub-gauge condition \cite{Chirilli:2015fza}. Then $A^\mathrm{phys}_\alpha (x) = A_\alpha (x)$. \eq{LGcomp1} becomes (after integration by parts)
\begin{align}\label{norm1}
L_G = \frac{1}{2 S^+ \, V^-} \int d^2 x \, d x^-  \bra{P, S} F^{+\alpha} (x) \, \left( {\un x} \times {\un \nabla} \right) \, A_\alpha (x) \, \ket{P, S}. 
\end{align}
Assuming that color trace ($2 \, \tr$) is implied in Eq.~(4) of \cite{Hatta:2016aoc}, and using $S^+ = P^+$ we write \eq{norm1} as
\begin{align}\label{norm2}
L_G = - \frac{1}{2 P^+ \, V^-} \int d^2 x \, d x^-  \bra{P, S} 2 \, \tr \left[ F^{+i} (x) \, \left( {\un x} \times {\un \nabla} \right) \, A^i (x) \right] \, \ket{P, S}. 
\end{align}
Observing that $E^i = - F^{0i} = - (F^{+i} + F^{-i})/\sqrt{2}$ and neglecting $F^{-i}$ in the infinite momentum frame (which we can do even for the sub-eikonal helicity-dependent gluon field, as it appears that helicity-dependent part of $F^{-i}$ is sub-sub-eikonal) we get $F^{+i} = - \sqrt{2} \, E^i$. On the other hand, $P^+ = \sqrt{2} \, E$, with $E$ the energy of the proton. We get
\begin{align}\label{norm3}
L_G = \frac{1}{2 E \, V^-} \int d^2 x \, d x^-  \bra{P, S} 2 \, \tr \left[ E^{i} (x) \, \left( {\un x} \times {\un \nabla} \right) \, A^i (x) \right] \, \ket{P, S}. 
\end{align}
This agrees with the first term in the second line of Eq.~(6.39) in \cite{Jaffe:1989jz}. Note that  ${\un \nabla}^i = \pd/\pd x^i = \pd_i = - \pd^i$.

The OAM definitions above in \eq{OAM0} (applied to gluons) and in Eq.~(29) of \cite{Hatta:2016aoc} (labeled HNXYZ) are very similar, and in fact would be identical if
\begin{align}\label{equality1}
\int d b^- \, P^+ \, \frac{1}{(2 \pi)^3} \, W^{here} (p,b) = W^{HNXYZ} (x, {\un p}, {\un b}).
\end{align}
Using the Wigner distributions from Eq.~(25) in \cite{Hatta:2016aoc} and \eq{Wig3} above we see that \eq{equality1} is satisfied if
\begin{align}\label{equality2}
& \int \frac{d^2 \Delta}{(2 \pi)^2} \, \bra{P+ \frac{{\un \Delta}}{2}, S} \, \tr \left[ F^{+i} \left({\un b} + \frac{z}{2} \right) \, {\cal U} \, F^{+i} \left({\un b} - \frac{z}{2} \right) \right] \, \ket{P- \frac{{\un \Delta}}{2}, S} \\ \notag & = 2 \, P^+ \, \int db^- \, \left\langle \tr \left[ F^{+i} \left(b + \frac{z}{2} \right) \, {\cal U} \, F^{+i} \left(b - \frac{z}{2} \right) \right]  \right\rangle ,
\end{align}
where we have replaced $\xi \to -z$ in \eq{Wig3} . The gauge link or links are denoted by a single $\cal U$ for brevity. Using \eq{matrix_el3} we see that \eq{equality2} is indeed correct since
\begin{align}
& 2 \, P^+ \, \int db^- \, \left\langle \tr \left[ F^{+i} \left(b + \frac{z}{2} \right) \, {\cal U} \, F^{+i} \left(b - \frac{z}{2} \right) \right]  \right\rangle \notag \\ & = 2 \, P^+ \, \int db^- \, \frac{1}{2 \, P^+} \, \int \frac{d^2 \Delta \, d \Delta^+}{(2 \pi)^3} \, e^{i b \cdot \Delta} \, \bra{P+ \frac{{\Delta}}{2}, S} \, \tr \left[ F^{+i} \left(\frac{z}{2} \right) \, {\cal U} \, F^{+i} \left( - \frac{z}{2} \right) \right] \, \ket{P- \frac{{\Delta}}{2}, S} \notag \\ & = \int \frac{d^2 \Delta}{(2 \pi)^2} \, e^{- i {\un b} \cdot {\un \Delta}} \, \bra{P+ \frac{{\un \Delta}}{2}, S} \, \tr \left[ F^{+i} \left( \frac{z}{2} \right) \, {\cal U} \, F^{+i} \left( - \frac{z}{2} \right) \right] \, \ket{P- \frac{{\un \Delta}}{2}, S} \notag \\ & = \int \frac{d^2 \Delta}{(2 \pi)^2} \, \bra{P+ \frac{{\un \Delta}}{2}, S} \, \tr \left[ F^{+i} \left({\un b} + \frac{z}{2} \right) \, {\cal U} \, F^{+i} \left({\un b} - \frac{z}{2} \right) \right] \, \ket{P- \frac{{\un \Delta}}{2}, S} . 
\end{align}
Hence our gluon OAM definition is equivalent to that in \cite{Jaffe:1989jz,Hatta:2016aoc}.


\section{Large nucleus limit}
\label{app:large}

Imagine the CGC limit, that is, consider the proton to be a large nucleus in the McLerran-Venugopalan (MV) model \cite{McLerran:1993ni,McLerran:1993ka,McLerran:1994vd} with one of the quarks in one of the nucleons polarized. In this case, at Born level, we can write using \eq{G50}
\begin{align}\label{CGC_Gi}
G_{10}^i = - \frac{\as^2 \, C_F}{N_c} \, \pi \, \epsilon^{ij} \, x^j_{10} \, \ln \frac{1}{x_{10} \Lambda} \, T (b),
\end{align} 
where $T(b)$ is the nuclear profile function and, as usual, ${\un b} = ({\un x}_0 + {\un x}_1)/2$. (See Sec.~4.2.1 of \cite{KovchegovLevin}, in particular Eq.~(4.32).) 

Inserting \eq{CGC_Gi} into \eq{OAM13} and observing that
\begin{align}
\int d^2 b \, b^j \, T(b) =0
\end{align}
due to the rotational symmetry of $T(b)$ we see that the first term in the right of \eq{OAM13} vanishes and one arrives at
\begin{align}\label{OAM1300}
\frac{d L_G}{d x} =  - \frac{1}{2} \, \Delta G (x,Q^2).
\end{align}

Below \eq{OAM15} we made a comment that for the proton target modeled to be a single quark the relation between the gluon helicity and OAM (at the same Born level) is
\begin{align}\label{OAM1500}
L_G (x, Q^2) =  -  \Delta G (x,Q^2),
\end{align} 
which is also in agreement with the parton model argument presented in Appendix~B of \cite{Hatta:2016aoc}. It appears that in the same Born (two-gluon exchange) approximation, the relation between the gluon helicity and OAM is different for the large-nucleus and single-quark targets, given by Eqs.~\eqref{OAM1300} and \eqref{OAM1500} respectively.

Interestingly, if we assume that (note the argument of $T$)
\begin{align}\label{CGC_Gi2}
G_{10}^i = - \frac{\as^2 \, C_F}{N_c} \, \pi \, \epsilon^{ij} \, x^j_{10} \, \ln \frac{1}{x_{10} \Lambda} \, T (x_1),
\end{align} 
or, in general that $G_{10}^i \sim T (x_1)$, then, employing \eq{OAM13} again (or \eq{OAM18}) we get
\begin{align}\label{OAM1600}
L_G (x, Q^2) =  0.
\end{align} 
It seems that slightly different assumptions about the argument of $T$, equivalent under the large-nucleus approximation of the MV model, give significantly different results for $L_G$. 

Indeed, we may vary the transverse position in the argument of the nuclear profile function by writing 
\begin{align}\label{CGC_Gi3}
G_{10}^i = - \frac{\as^2 \, C_F}{N_c} \, \pi \, \epsilon^{ij} \, x^j_{10} \, \ln \frac{1}{x_{10} \Lambda} \, T [\xi \, {\un x}_1 + (1-\xi) \, {\un x}_0]
\end{align} 
with $\xi$ a real dimensionless number. Then repeating the above steps used in arriving at \eq{OAM18} and \eq{OAM1600} we get
\begin{align}
L_G (x, Q^2) = (\xi -1) \, \Delta G (x,Q^2). 
\end{align}
Our result \eqref{LGDG2} above corresponds to 
\begin{align}\label{xif}
\xi  = 1 + \frac{\alpha_h^q}{4} \, \ln \frac{Q^2}{\Lambda^2} ,
\end{align}
that is, to $\xi$ slightly larger than 1. Note that $\xi \approx 1$ gives $T({\un x}_1)$ in \eq{CGC_Gi3}, indicating that the position of the polarized (anti)quark is ``more important". This is consistent with other findings in this work. 

In the framework of the simple toy model for the polarized amplitude in \eq{CGC_Gi3}, our conclusion \eqref{LGDG2} in the main text appears to imply that the gluon OAM begins with $\xi =0$ in the initial conditions (at the Born level) corresponding to $L_G = - \Delta G$, and then, via the DLA evolution, this parameter $\xi$ evolves to $\xi \gsim 1$, as given by \eq{xif}, with the relation \eqref{LGDG2} between the gluon OAM and helicity. The physical reason between such a ``center-of-mass" shift is not clear at present.



\begin{thebibliography}{10}

\bibitem{Kovchegov:2015pbl}
Y.~V. Kovchegov, D.~Pitonyak and M.~D. Sievert, \emph{{Helicity Evolution at
  Small-x}}, \href{https://doi.org/10.1007/JHEP01(2016)072}{\emph{JHEP}
  {\bfseries 01} (2016) 072},
  [\href{https://arxiv.org/abs/1511.06737}{{\ttfamily 1511.06737}}].

\bibitem{Kovchegov:2016weo}
Y.~V. Kovchegov, D.~Pitonyak and M.~D. Sievert, \emph{{Small-$x$ asymptotics of
  the quark helicity distribution}},
  \href{https://doi.org/10.1103/PhysRevLett.118.052001}{\emph{Phys. Rev. Lett.}
  {\bfseries 118} (2017) 052001},
  [\href{https://arxiv.org/abs/1610.06188}{{\ttfamily 1610.06188}}].

\bibitem{Kovchegov:2016zex}
Y.~V. Kovchegov, D.~Pitonyak and M.~D. Sievert, \emph{{Helicity Evolution at
  Small $x$: Flavor Singlet and Non-Singlet Observables}},
  \href{https://doi.org/10.1103/PhysRevD.95.014033}{\emph{Phys. Rev.}
  {\bfseries D95} (2017) 014033},
  [\href{https://arxiv.org/abs/1610.06197}{{\ttfamily 1610.06197}}].

\bibitem{Kovchegov:2017jxc}
Y.~V. Kovchegov, D.~Pitonyak and M.~D. Sievert, \emph{{Small-$x$ Asymptotics of
  the Quark Helicity Distribution: Analytic Results}},
  \href{https://arxiv.org/abs/1703.05809}{{\ttfamily 1703.05809}}.

\bibitem{Kovchegov:2017lsr}
Y.~V. Kovchegov, D.~Pitonyak and M.~D. Sievert, \emph{{Small-$x$ Asymptotics of
  the Gluon Helicity Distribution}},
  \href{https://doi.org/10.1007/JHEP10(2017)198}{\emph{JHEP} {\bfseries 10}
  (2017) 198}, [\href{https://arxiv.org/abs/1706.04236}{{\ttfamily
  1706.04236}}].

\bibitem{Kovchegov:2018znm}
Y.~V. Kovchegov and M.~D. Sievert, \emph{{Small-$x$ Helicity Evolution: an
  Operator Treatment}},  \href{https://arxiv.org/abs/1808.09010}{{\ttfamily
  1808.09010}}.

\bibitem{Jaffe:1989jz}
R.~L. Jaffe and A.~Manohar, \emph{{The G(1) Problem: Fact and Fantasy on the
  Spin of the Proton}},
  \href{https://doi.org/10.1016/0550-3213(90)90506-9}{\emph{Nucl. Phys.}
  {\bfseries B337} (1990) 509--546}.

\bibitem{Ji:1996ek}
X.-D. Ji, \emph{{Gauge-Invariant Decomposition of Nucleon Spin}},
  \href{https://doi.org/10.1103/PhysRevLett.78.610}{\emph{Phys. Rev. Lett.}
  {\bfseries 78} (1997) 610--613},
  [\href{https://arxiv.org/abs/hep-ph/9603249}{{\ttfamily hep-ph/9603249}}].

\bibitem{Ji:2012sj}
X.~Ji, X.~Xiong and F.~Yuan, \emph{{Proton Spin Structure from Measurable
  Parton Distributions}},
  \href{https://doi.org/10.1103/PhysRevLett.109.152005}{\emph{Phys. Rev. Lett.}
  {\bfseries 109} (2012) 152005},
  [\href{https://arxiv.org/abs/1202.2843}{{\ttfamily 1202.2843}}].

\bibitem{Leader:2013jra}
E.~Leader and C.~Lorce, \emph{{The angular momentum controversy: What’s it
  all about and does it matter?}},
  \href{https://doi.org/10.1016/j.physrep.2014.02.010}{\emph{Phys.Rept.}
  {\bfseries 541} (2014) 163--248},
  [\href{https://arxiv.org/abs/1309.4235}{{\ttfamily 1309.4235}}].

\bibitem{Bashinsky:1998if}
S.~Bashinsky and R.~L. Jaffe, \emph{{Quark and gluon orbital angular momentum
  and spin in hard processes}},
  \href{https://doi.org/10.1016/S0550-3213(98)00559-8}{\emph{Nucl. Phys.}
  {\bfseries B536} (1998) 303--317},
  [\href{https://arxiv.org/abs/hep-ph/9804397}{{\ttfamily hep-ph/9804397}}].

\bibitem{Hagler:1998kg}
P.~Hagler and A.~Schafer, \emph{{Evolution equations for higher moments of
  angular momentum distributions}},
  \href{https://doi.org/10.1016/S0370-2693(98)00414-6}{\emph{Phys. Lett.}
  {\bfseries B430} (1998) 179--185},
  [\href{https://arxiv.org/abs/hep-ph/9802362}{{\ttfamily hep-ph/9802362}}].

\bibitem{Harindranath:1998ve}
A.~Harindranath and R.~Kundu, \emph{{On Orbital angular momentum in deep
  inelastic scattering}},
  \href{https://doi.org/10.1103/PhysRevD.59.116013}{\emph{Phys. Rev.}
  {\bfseries D59} (1999) 116013},
  [\href{https://arxiv.org/abs/hep-ph/9802406}{{\ttfamily hep-ph/9802406}}].

\bibitem{Hatta:2012cs}
Y.~Hatta and S.~Yoshida, \emph{{Twist analysis of the nucleon spin in QCD}},
  \href{https://doi.org/10.1007/JHEP10(2012)080}{\emph{JHEP} {\bfseries 10}
  (2012) 080}, [\href{https://arxiv.org/abs/1207.5332}{{\ttfamily 1207.5332}}].

\bibitem{Ji:2012ba}
X.~Ji, X.~Xiong and F.~Yuan, \emph{{Probing Parton Orbital Angular Momentum in
  Longitudinally Polarized Nucleon}},
  \href{https://doi.org/10.1103/PhysRevD.88.014041}{\emph{Phys. Rev.}
  {\bfseries D88} (2013) 014041},
  [\href{https://arxiv.org/abs/1207.5221}{{\ttfamily 1207.5221}}].

\bibitem{Hatta:2018itc}
Y.~Hatta and D.-J. Yang, \emph{{On the small-$x$ behavior of the orbital
  angular momentum distributions in QCD}},
  \href{https://doi.org/10.1016/j.physletb.2018.03.081}{\emph{Phys. Lett.}
  {\bfseries B781} (2018) 213--219},
  [\href{https://arxiv.org/abs/1802.02716}{{\ttfamily 1802.02716}}].

\bibitem{Martin:1998fe}
O.~Martin, P.~Hagler and A.~Schafer, \emph{{Numerical solution of the evolution
  equation for orbital angular momentum of partons in the nucleon}},
  \href{https://doi.org/10.1016/S0370-2693(99)00025-8}{\emph{Phys. Lett.}
  {\bfseries B448} (1999) 99--106},
  [\href{https://arxiv.org/abs/hep-ph/9810474}{{\ttfamily hep-ph/9810474}}].

\bibitem{Hatta:2016aoc}
Y.~Hatta, Y.~Nakagawa, F.~Yuan, Y.~Zhao and B.~Xiao, \emph{{Gluon orbital
  angular momentum at small-$x$}},
  \href{https://doi.org/10.1103/PhysRevD.95.114032}{\emph{Phys. Rev.}
  {\bfseries D95} (2017) 114032},
  [\href{https://arxiv.org/abs/1612.02445}{{\ttfamily 1612.02445}}].

\bibitem{Kuraev:1977fs}
E.~A. Kuraev, L.~N. Lipatov and V.~S. Fadin, \emph{{The Pomeranchuk
  singlularity in non-Abelian gauge theories}}, {\emph{Sov. Phys. JETP}
  {\bfseries 45} (1977) 199--204}.

\bibitem{Balitsky:1978ic}
I.~Balitsky and L.~Lipatov, \emph{{The Pomeranchuk Singularity in Quantum
  Chromodynamics}}, {\emph{Sov.J.Nucl.Phys.} {\bfseries 28} (1978) 822--829}.

\bibitem{Balitsky:1995ub}
I.~Balitsky, \emph{{Operator expansion for high-energy scattering}},
  \href{https://doi.org/10.1016/0550-3213(95)00638-9}{\emph{Nucl. Phys.}
  {\bfseries B463} (1996) 99--160},
  [\href{https://arxiv.org/abs/hep-ph/9509348}{{\ttfamily hep-ph/9509348}}].

\bibitem{Balitsky:1998ya}
I.~Balitsky, \emph{Factorization and high-energy effective action},
  {\emph{Phys. Rev.} {\bfseries D60} (1999) 014020},
  [\href{https://arxiv.org/abs/hep-ph/9812311}{{\ttfamily hep-ph/9812311}}].

\bibitem{Kovchegov:1999yj}
Y.~V. Kovchegov, \emph{Small-x {$F_2$} structure function of a nucleus
  including multiple pomeron exchanges}, {\emph{Phys. Rev.} {\bfseries D60}
  (1999) 034008}, [\href{https://arxiv.org/abs/hep-ph/9901281}{{\ttfamily
  hep-ph/9901281}}].

\bibitem{Kovchegov:1999ua}
Y.~V. Kovchegov, \emph{Unitarization of the {BFKL} pomeron on a nucleus},
  {\emph{Phys. Rev.} {\bfseries D61} (2000) 074018},
  [\href{https://arxiv.org/abs/hep-ph/9905214}{{\ttfamily hep-ph/9905214}}].

\bibitem{Jalilian-Marian:1997dw}
J.~Jalilian-Marian, A.~Kovner and H.~Weigert, \emph{The {Wilson}
  renormalization group for low x physics: Gluon evolution at finite parton
  density}, {\emph{Phys. Rev.} {\bfseries D59} (1998) 014015},
  [\href{https://arxiv.org/abs/hep-ph/9709432}{{\ttfamily hep-ph/9709432}}].

\bibitem{Jalilian-Marian:1997gr}
J.~Jalilian-Marian, A.~Kovner, A.~Leonidov and H.~Weigert, \emph{The {Wilson}
  renormalization group for low x physics: Towards the high density regime},
  {\emph{Phys. Rev.} {\bfseries D59} (1998) 014014},
  [\href{https://arxiv.org/abs/hep-ph/9706377}{{\ttfamily hep-ph/9706377}}].

\bibitem{Iancu:2001ad}
E.~Iancu, A.~Leonidov and L.~D. McLerran, \emph{{The renormalization group
  equation for the color glass condensate}},
  \href{https://doi.org/10.1016/S0370-2693(01)00524-X}{\emph{Phys. Lett.}
  {\bfseries B510} (2001) 133--144}.

\bibitem{Iancu:2000hn}
E.~Iancu, A.~Leonidov and L.~D. McLerran, \emph{Nonlinear gluon evolution in
  the color glass condensate. {I}}, {\emph{Nucl. Phys.} {\bfseries A692} (2001)
  583--645}, [\href{https://arxiv.org/abs/hep-ph/0011241}{{\ttfamily
  hep-ph/0011241}}].

\bibitem{Kirschner:1983di}
R.~Kirschner and L.~Lipatov, \emph{{Double Logarithmic Asymptotics and Regge
  Singularities of Quark Amplitudes with Flavor Exchange}},
  \href{https://doi.org/10.1016/0550-3213(83)90178-5}{\emph{Nucl.Phys.}
  {\bfseries B213} (1983) 122--148}.

\bibitem{Kirschner:1985cb}
R.~Kirschner, \emph{{Regge Asymptotics of Scattering Amplitudes in the
  Logarithmic Approximation of {QCD}}},
  \href{https://doi.org/10.1007/BF01559604}{\emph{Z. Phys.} {\bfseries C31}
  (1986) 135}.

\bibitem{Kirschner:1994vc}
R.~Kirschner, \emph{{Regge asymptotics of scattering with flavor exchange in
  QCD}}, \href{https://doi.org/10.1007/BF01624588}{\emph{Z.Phys.} {\bfseries
  C67} (1995) 459--466},
  [\href{https://arxiv.org/abs/hep-th/9404158}{{\ttfamily hep-th/9404158}}].

\bibitem{Kirschner:1994rq}
R.~Kirschner, \emph{{Reggeon interactions in perturbative QCD}},
  \href{https://doi.org/10.1007/BF01556138}{\emph{Z.Phys.} {\bfseries C65}
  (1995) 505--510}, [\href{https://arxiv.org/abs/hep-th/9407085}{{\ttfamily
  hep-th/9407085}}].

\bibitem{Griffiths:1999dj}
S.~Griffiths and D.~Ross, \emph{{Studying the perturbative Reggeon}},
  \href{https://doi.org/10.1007/s100529900240}{\emph{Eur.Phys.J.} {\bfseries
  C12} (2000) 277--286},
  [\href{https://arxiv.org/abs/hep-ph/9906550}{{\ttfamily hep-ph/9906550}}].

\bibitem{Itakura:2003jp}
K.~Itakura, Y.~V. Kovchegov, L.~McLerran and D.~Teaney, \emph{{Baryon stopping
  and valence quark distribution at small x}},
  \href{https://doi.org/10.1016/j.nuclphysa.2003.10.016}{\emph{Nucl. Phys.}
  {\bfseries A730} (2004) 160--190},
  [\href{https://arxiv.org/abs/hep-ph/0305332}{{\ttfamily hep-ph/0305332}}].

\bibitem{Bartels:2003dj}
J.~Bartels and M.~Lublinsky, \emph{{Quark anti-quark exchange in gamma* gamma*
  scattering}},
  \href{https://doi.org/10.1088/1126-6708/2003/09/076}{\emph{JHEP} {\bfseries
  0309} (2003) 076}, [\href{https://arxiv.org/abs/hep-ph/0308181}{{\ttfamily
  hep-ph/0308181}}].

\bibitem{Bartels:1995iu}
J.~Bartels, B.~Ermolaev and M.~Ryskin, \emph{{Nonsinglet contributions to the
  structure function g1 at small x}}, {\emph{Z.Phys.} {\bfseries C70} (1996)
  273--280}, [\href{https://arxiv.org/abs/hep-ph/9507271}{{\ttfamily
  hep-ph/9507271}}].

\bibitem{Bartels:1996wc}
J.~Bartels, B.~Ermolaev and M.~Ryskin, \emph{{Flavor singlet contribution to
  the structure function G(1) at small x}},
  \href{https://doi.org/10.1007/s002880050285}{\emph{Z.Phys.} {\bfseries C72}
  (1996) 627--635}, [\href{https://arxiv.org/abs/hep-ph/9603204}{{\ttfamily
  hep-ph/9603204}}].

\bibitem{Belitsky:2003nz}
A.~V. Belitsky, X.-d. Ji and F.~Yuan, \emph{{Quark imaging in the proton via
  quantum phase space distributions}},
  \href{https://doi.org/10.1103/PhysRevD.69.074014}{\emph{Phys. Rev.}
  {\bfseries D69} (2004) 074014},
  [\href{https://arxiv.org/abs/hep-ph/0307383}{{\ttfamily hep-ph/0307383}}].

\bibitem{Lorce:2011kd}
C.~Lorce and B.~Pasquini, \emph{{Quark Wigner Distributions and Orbital Angular
  Momentum}}, \href{https://doi.org/10.1103/PhysRevD.84.014015}{\emph{Phys.
  Rev.} {\bfseries D84} (2011) 014015},
  [\href{https://arxiv.org/abs/1106.0139}{{\ttfamily 1106.0139}}].

\bibitem{Hatta:2011ku}
Y.~Hatta, \emph{{Notes on the orbital angular momentum of quarks in the
  nucleon}}, \href{https://doi.org/10.1016/j.physletb.2012.01.024}{\emph{Phys.
  Lett.} {\bfseries B708} (2012) 186--190},
  [\href{https://arxiv.org/abs/1111.3547}{{\ttfamily 1111.3547}}].

\bibitem{Lorce:2011ni}
C.~Lorce, B.~Pasquini, X.~Xiong and F.~Yuan, \emph{{The quark orbital angular
  momentum from Wigner distributions and light-cone wave functions}},
  \href{https://doi.org/10.1103/PhysRevD.85.114006}{\emph{Phys. Rev.}
  {\bfseries D85} (2012) 114006},
  [\href{https://arxiv.org/abs/1111.4827}{{\ttfamily 1111.4827}}].

\bibitem{Mulders:1995dh}
P.~J. Mulders and R.~D. Tangerman, \emph{{The Complete tree level result up to
  order 1/Q for polarized deep inelastic leptoproduction}},
  \href{https://doi.org/10.1016/0550-3213(95)00632-X}{\emph{Nucl. Phys.}
  {\bfseries B461} (1996) 197--237},
  [\href{https://arxiv.org/abs/hep-ph/9510301}{{\ttfamily hep-ph/9510301}}].

\bibitem{Gribov:1984tu}
L.~V. Gribov, E.~M. Levin and M.~G. Ryskin, \emph{{Semihard Processes in QCD}},
  {\emph{Phys. Rept.} {\bfseries 100} (1983) 1--150}.

\bibitem{Iancu:2003xm}
E.~Iancu and R.~Venugopalan, \emph{The color glass condensate and high energy
  scattering in {QCD}},  \href{https://arxiv.org/abs/hep-ph/0303204}{{\ttfamily
  hep-ph/0303204}}.

\bibitem{Weigert:2005us}
H.~Weigert, \emph{Evolution at small {$x_{bj}$: The Color Glass Condensate}},
  {\emph{Prog. Part. Nucl. Phys.} {\bfseries 55} (2005) 461--565},
  [\href{https://arxiv.org/abs/hep-ph/0501087}{{\ttfamily hep-ph/0501087}}].

\bibitem{Jalilian-Marian:2005jf}
J.~Jalilian-Marian and Y.~V. Kovchegov, \emph{Saturation physics and deuteron
  gold collisions at {RHIC}}, {\emph{Prog. Part. Nucl. Phys.} {\bfseries 56}
  (2006) 104--231}, [\href{https://arxiv.org/abs/hep-ph/0505052}{{\ttfamily
  hep-ph/0505052}}].

\bibitem{Gelis:2010nm}
F.~Gelis, E.~Iancu, J.~Jalilian-Marian and R.~Venugopalan, \emph{{The Color
  Glass Condensate}},
  \href{https://doi.org/10.1146/annurev.nucl.010909.083629}{\emph{Ann.Rev.Nucl.Part.Sci.}
  {\bfseries 60} (2010) 463--489},
  [\href{https://arxiv.org/abs/1002.0333}{{\ttfamily 1002.0333}}].

\bibitem{Albacete:2014fwa}
J.~L. Albacete and C.~Marquet, \emph{{Gluon saturation and initial conditions
  for relativistic heavy ion collisions}},
  \href{https://doi.org/10.1016/j.ppnp.2014.01.004}{\emph{Prog.Part.Nucl.Phys.}
  {\bfseries 76} (2014) 1--42},
  [\href{https://arxiv.org/abs/1401.4866}{{\ttfamily 1401.4866}}].

\bibitem{KovchegovLevin}
Y.~V. Kovchegov and E.~Levin, \emph{Quantum Chromodynamics at High Energy}.
\newblock Cambridge University Press, 2012.

\bibitem{Lepage:1980fj}
G.~P. Lepage and S.~J. Brodsky, \emph{Exclusive processes in perturbative
  quantum chromodynamics}, {\emph{Phys. Rev.} {\bfseries D22} (1980) 2157}.

\bibitem{Kovchegov:2018zeq}
Y.~V. Kovchegov and M.~D. Sievert, \emph{{Valence Quark Transversity at Small
  $x$}},  \href{https://arxiv.org/abs/1808.10354}{{\ttfamily 1808.10354}}.

\bibitem{Mueller:2012bn}
A.~Mueller and S.~Munier, \emph{{$p_{\perp}$-broadening and production
  processes versus dipole/quadrupole amplitudes at next-to-leading order}},
  \href{https://doi.org/10.1016/j.nuclphysa.2012.08.005}{\emph{Nucl.Phys.}
  {\bfseries A893} (2012) 43--86},
  [\href{https://arxiv.org/abs/1206.1333}{{\ttfamily 1206.1333}}].

\bibitem{Dominguez:2011wm}
F.~Dominguez, C.~Marquet, B.-W. Xiao and F.~Yuan, \emph{{Universality of
  Unintegrated Gluon Distributions at small x}},
  \href{https://doi.org/10.1103/PhysRevD.83.105005}{\emph{Phys.Rev.} {\bfseries
  D83} (2011) 105005}, [\href{https://arxiv.org/abs/1101.0715}{{\ttfamily
  1101.0715}}].

\bibitem{Chirilli:2015fza}
G.~A. Chirilli, Y.~V. Kovchegov and D.~E. Wertepny, \emph{{Regularization of
  the Light-Cone Gauge Gluon Propagator Singularities Using Sub-Gauge
  Conditions}}, \href{https://doi.org/10.1007/JHEP12(2015)138}{\emph{JHEP}
  {\bfseries 12} (2015) 138},
  [\href{https://arxiv.org/abs/1508.07962}{{\ttfamily 1508.07962}}].

\bibitem{McLerran:1993ni}
L.~D. McLerran and R.~Venugopalan, \emph{Computing quark and gluon distribution
  functions for very large nuclei}, {\emph{Phys. Rev.} {\bfseries D49} (1994)
  2233--2241}, [\href{https://arxiv.org/abs/hep-ph/9309289}{{\ttfamily
  hep-ph/9309289}}].

\bibitem{McLerran:1993ka}
L.~D. McLerran and R.~Venugopalan, \emph{Gluon distribution functions for very
  large nuclei at small transverse momentum}, {\emph{Phys. Rev.} {\bfseries
  D49} (1994) 3352--3355},
  [\href{https://arxiv.org/abs/hep-ph/9311205}{{\ttfamily hep-ph/9311205}}].

\bibitem{McLerran:1994vd}
L.~D. McLerran and R.~Venugopalan, \emph{Green's functions in the color field
  of a large nucleus}, {\emph{Phys. Rev.} {\bfseries D50} (1994) 2225--2233},
  [\href{https://arxiv.org/abs/hep-ph/9402335}{{\ttfamily hep-ph/9402335}}].

\end{thebibliography}

\providecommand{\href}[2]{#2}\begingroup\raggedright\endgroup

\end{document}